\pdfoutput=1
\documentclass[letterpaper,journal]{IEEEtran}
\usepackage{amsmath,amsfonts,amssymb}
\usepackage{algorithmic}
\usepackage{algorithm}
\usepackage{array}
\usepackage[caption=false,font=normalsize,labelfont=sf,textfont=sf]{subfig}
\usepackage{textcomp}
\usepackage{stfloats}
\usepackage{url}
\usepackage{verbatim}
\usepackage{graphicx}
\usepackage{cite}

\usepackage{booktabs}
\usepackage{tabularx}
\usepackage{threeparttable}

\usepackage{physics}
\usepackage{glossaries}
\glsdisablehyper 
\usepackage{xcolor}
\usepackage{ulem}
\usepackage{hyperref}

\newacronym{qkd}{QKD}{Quantum Key Distribution}
\newacronym{wdm}{WDM}{Wavelength Division Multiplexing}
\newacronym{qnlse}{QNLSE}{Quantum
Nonlinear Schrödinger equation}
\newacronym{rrc}{RRC}{root-raised cosine}
\newacronym{qam}{QAM}{Quadrature Amplitude Modulation}
\newacronym{itu}{ITU}{International Telecommunication Union}

\begin{document}

\title{Kerr-Induced Impairments in Quantum–Classical Fiber Coexistence: A Quantum-Stochastic Analysis 
}

\author{Sumit Chaudhary, Shahram Dehdashti, Igor Litvin, Janis N\"otzel \\ ~\IEEEmembership{}
\IEEEauthorblockA{\textit{Emmy-Noether Group Theoretical Quantum Systems Design} \\
\textit{Technical University of Munich}\\
Munich, Germany \\
\{sumit.chaudhary, shahram.dehdashti, igor.litvin, janis.noetzel\}@tum.de }

\thanks{This paper was produced by the IEEE Publication Technology Group. They are in Piscataway, NJ.}
\thanks{Manuscript received April 19, 2021; revised August 16, 2021.}}

\markboth{Journal of \LaTeX\ Class Files,~Vol.~14, No.~8, August~2021}%
{Shell \MakeLowercase{\textit{et al.}}: A Sample Article Using IEEEtran.cls for IEEE Journals}


\maketitle

\begin{abstract}
Quantum and classical optical signals sharing a fiber can interact through Kerr cross-phase modulation without interchannel photon transfer, potentially degrading phase- and mode-sensitive quantum communication even when direct spectral leakage is negligible. We develop a two-channel quantum-stochastic propagation framework based on the positive-$P$ representation for a weak coherent pulse train co-propagating with a strong, quadrature-amplitude-modulated classical signal. Both fields are treated quantum mechanically, incorporating self- and cross-phase modulation, chromatic dispersion, group-velocity mismatch, attenuation, and Kerr-induced quantum diffusion. We characterize the classical-channel influence using three complementary observables: the frequency-resolved quadrature-fluctuation spectrum, complex temporal-mode overlap, and adjacent-pulse interference visibility. Simulations of a $100$-km standard-fiber link reveal markedly different responses across these observables. As classical launch power increases, the extremal quadrature-fluctuation spectra remain close to the shot-noise level, whereas the temporal-mode and visibility penalties reach approximately $4.5\%$ and $6.0\%$, respectively, at $100$ mW. Applying a $1\%$ mean-penalty criterion gives launch-power thresholds of approximately $46$ mW for temporal-mode overlap and $40$ mW for visibility, yielding a joint Kerr-limited threshold near $40$ mW for the investigated configuration. These results demonstrate that temporal-mode deformation and reduced pulse-to-pulse coherence can occur while the evaluated quadrature-fluctuation spectra remain near shot noise, establishing a complementary criterion for selecting classical launch power in quantum–classical fiber co-propagation.

\end{abstract}

\begin{IEEEkeywords}
Cross-phase modulation, positive-P representation, quantum–classical coexistence, quantum communication
\end{IEEEkeywords}

\section{Introduction}
\label{sec:introduction}

\IEEEPARstart{S}{haring}  installed optical fiber is one of the most cost-effective routes to scalable quantum networking. It does, however, place weak coherent pulses—typically containing fewer than one photon on average—alongside classical data channels operating at milliwatt-scale launch powers. Most quantum–classical coexistence budgets therefore focus on photons scattered or leaked into the quantum receiver through Raman scattering, amplified spontaneous emission, finite filter isolation, Rayleigh scattering, or four-wave mixing. Kerr cross-phase modulation creates a fundamentally different failure mode: without directly transferring energy into the quantum-channel bandwidth, it can distort the temporal mode and relative phase on which a quantum receiver relies. Consequently, a link may appear acceptable according to a conventional in-band photon-noise budget while already exhibiting degradation of its mode matching or phase coherence.

Quantum--classical coexistence has been demonstrated for discrete-variable, continuous-variable, and measurement-device-independent QKD over diverse wavelength allocations, modulation formats, and transmission distances \cite{peters2009dense,eraerds2010quantum,patel2012coexistence,kumar2015coexistence,wang2017long,valivarthi2019measurement}. Recent experiments have extended shared-fiber operation to densely populated WDM systems, terabit-per-second traffic, bidirectional links, and launch powers representative of deployed networks \cite{eriksson2019wavelength,honz2023first,gavignet2023co,wang2024time,dou2024coexistence}. Recent demonstrations have further broadened the operating regime of quantum–classical coexistence. Chapman et al. distributed two-mode squeezing over deployed fiber while frequency-multiplexing the quantum modes with conventional network signals \cite{chapman2023two}. Hajomer et al. demonstrated continuous-variable QKD coexisting with a fully populated CWDM system over 120 km \cite{hajomer2025coexistence}, while Linale et al. reported multi-day C-band operation of polarization-encoded QKD with classical traffic over deployed standard fiber \cite{linale2026coexistence}. Zischler et al. developed a classical-field model of coexistence based on interference power spectral density \cite{zischler2026accurate}, while XPM-induced differential phase fluctuations can additionally degrade visibility without interchannel power transfer. 

In parallel, positive-\(P\) quantum-stochastic generalized nonlinear Schrödinger equations have been applied to investigate quantum noise in nonlinear fiber propagation, including supercontinuum generation \cite{safaei2022quantum}. These developments underscore the practical importance of shared-fiber quantum networking and motivate channel-resolved models that connect Kerr-mediated quantum fluctuations to temporal-mode deformation and pulse-to-pulse coherence.

Unlike scattering and leakage mechanisms, the instantaneous Kerr interaction is number conserving and does not require direct energy transfer into the quantum-channel bandwidth. Through cross-phase modulation (XPM), intensity variations in a data-modulated classical waveform impose a time-dependent phase on the quantum field. Chromatic dispersion and group-velocity mismatch can subsequently convert this phase modulation into pulse-dependent temporal deformation and relative phase evolution. Such impairment may therefore remain inconspicuous in a conventional photon-counting noise budget while degrading temporal-mode matching and the coherence required by phase- and time-bin-sensitive receivers.

Deterministic coupled-field models can describe XPM-induced phase modulation and pulse reshaping but cannot, by themselves, provide the normally ordered quantum correlations required for shot-noise-referenced quadrature spectra. The quantum nonlinear Schrödinger equation, expressed through the positive-$P$ representation, maps the operator evolution onto stochastic equations whose ensemble moments reproduce normally ordered observables \cite{drummond1987quantum,b1,drummond2001quantum,drummond1980generalised}. This formulation enables the weak quantum signal and high-power classical waveform to be propagated within a unified quantum-mechanical model rather than representing the classical channel as a prescribed nonlinear phase.

In this work, we derive a two-channel positive-$P$ model for a weak coherent pulse train co-propagating with a \gls{rrc}-shaped $M$-QAM signal. Starting from a two-field Kerr Hamiltonian, the model jointly incorporates SPM, XPM, second-order dispersion, group-velocity mismatch, attenuation, and Kerr-induced quantum diffusion; the full derivation is provided in the Supplementary Material. We evaluate three complementary observables: the quadrature-fluctuation spectrum, complex temporal-mode overlap, and adjacent-pulse interference visibility, which respectively probe quantum fluctuations, mean-envelope deformation, and pulse-to-pulse coherence. 
We use the dark-fiber case, in which only the quantum signal propagates through the fiber, as the ideal reference for quantifying classical-channel-induced temporal-mode deformation.

For a 100-km standard single-mode fiber link, increasing the classical launch power leaves the shot-noise-normalized quadrature spectra close to the unit shot-noise level but produce clear temporal-mode and visibility penalties. At 100~mW, the normalized overlap with the dark-fiber reference is $F\simeq0.955$ and the mean visibility is $V\simeq0.94$, corresponding to penalties $1-F\simeq4.5\%$ and $1-V\simeq6.0\%$, respectively. Their 1\% thresholds occur near 46 and 40 mW, yielding a joint coexistence threshold of approximately 40 mW for the investigated configuration. The results demonstrate that mode deformation and coherence loss can precede appreciable quadrature-noise growth and provide a metric-based launch-power criterion for designing quantum--classical co-propagation networks.

The remainder of this paper is organized as follows. Section~II introduces the two-channel quantum-stochastic propagation model. Section~III defines the quantum and classical input signals, the three performance metrics, and the numerical procedure. Section~IV presents the power- and distance-dependent propagation results. Section~V discusses their physical interpretation, implications for quantum communication, and scope of the model. Section~VI concludes the paper.

\section{Quantum-Stochastic Model of Co-Propagation}
\label{sec:propagation-model}

\subsection{Physical system and assumptions}
\label{subsec:physical-system}

\begin{figure*}
    \centering
    \includegraphics[width=0.98\linewidth]{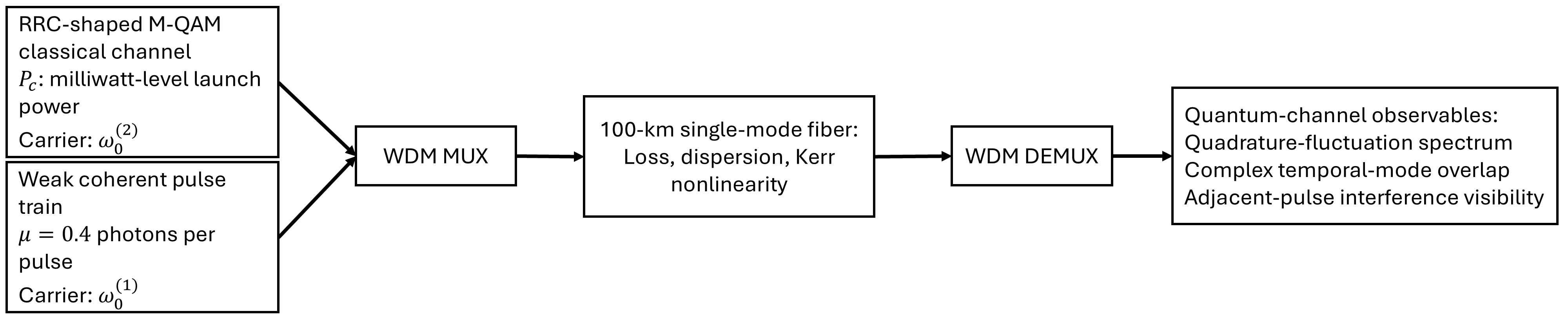}
    \caption{Schematic of the investigated quantum–classical co-propagation system. A weak coherent pulse train at carrier frequency $\omega_1$ is wavelength-multiplexed with an \gls{rrc}-shaped $M$-QAM classical channel at $\omega_2$. The two channels co-propagate through a 100-km single-mode fiber subject to attenuation, chromatic dispersion, group-velocity mismatch, self-phase modulation, and cross-phase modulation. After wavelength demultiplexing, the quantum channel is characterized using its quadrature-fluctuation spectrum, complex temporal-mode overlap, and adjacent-pulse interference visibility.}
    \label{fig:illustration}
\end{figure*}

We consider the quantum–classical co-propagation system illustrated in Fig. \ref{fig:illustration}. A weak coherent pulse train centered at angular frequency $\omega_0^{(1)}$ and an \gls{rrc}-shaped $M$-QAM classical waveform centered at $\omega_0^{(2)}$ are combined using a wavelength-division multiplexer and launched into a 100-km single-mode fiber. The two channels experience linear attenuation, chromatic dispersion, group-velocity mismatch, self-phase modulation, and cross-phase modulation. After wavelength demultiplexing, the quantum channel is characterized using its quadrature-fluctuation spectrum, complex temporal-mode overlap, and adjacent-pulse interference visibility. In the following, channel 1 denotes the quantum channel and channel 2 denotes the classical data channel.
The two fields have group velocities $v_1$ and $v_2$ and group-velocity-dispersion coefficients
\begin{equation}
k''^{(j)} =
\left.
\frac{\partial^2 k}{\partial\omega^2}
\right|_{\omega_0^{(j)}},
\qquad j\in \{1,2\}.
\label{eq:gvd-definition}
\end{equation}
Although the channels differ substantially in their occupation numbers, both optical fields are quantized and evolved within the same theoretical framework.

The derivation begins by expanding each field in longitudinal guided modes, following the standard quantization of propagating fields in nonlinear optical fibers~\cite{drummond1987quantum}. The corresponding annihilation operators $\hat{a}_m$ and $\hat{b}_m$ satisfy
\begin{equation}
[\hat{a}_m,\hat{a}_{m'}^\dagger]
=
[\hat{b}_m,\hat{b}_{m'}^\dagger]
=
\delta_{mm'},
\qquad
[\hat{a}_m,\hat{b}_{m'}^\dagger]=0.
\label{eq:channel-commutators}
\end{equation}
The fiber Hamiltonian \cite{drummond2001quantum, b1} contains the linear dispersive energy of the two fields, an instantaneous third-order nonlinear interaction, and independent reservoirs describing attenuation in the two channels. After applying the rotating-wave approximation, the number-conserving nonlinear terms consist of self-phase modulation (SPM) within each channel and cross-phase modulation (XPM) between the channels. Rapidly oscillating interchannel four-wave-mixing terms are not retained. The complete Hamiltonian construction and its reduction to a localized two-channel model are given in the Supplementary Material.

The model assumes a scalar, polarization-preserving guided mode for each carrier and identical polarization and transverse-mode overlap for the two channels. Dispersion is retained through second order, the electronic Kerr response is treated as instantaneous, and linear loss is represented by independent Markovian reservoirs. The derivation further employs the unidirectional slowly varying-envelope approximation. Delayed Raman and Brillouin responses, polarization-mode coupling, higher-order dispersion, and self-steepening are outside the present model.

\subsection{Positive-\texorpdfstring{$P$}{P} representation and scaling}
\label{subsec:positive-p-scaling}

After tracing out the reservoir degrees of freedom, we use
$\hat{\rho}=\operatorname{Tr}_{\mathrm{R}} (\hat{\rho}_{\mathrm{tot}})$ to denote the reduced density operator of the two optical fields. The positive-$P$ representation expands this operator over a doubled complex phase space \cite{drummond1980generalised,b1,drummond2001quantum}:

\begin{align}
\hat{\rho}
&=
\int
P(\boldsymbol{\lambda})
\hat{\Lambda}(\boldsymbol{\lambda})
\,\mathrm{d}\mu(\boldsymbol{\lambda}),
\qquad
\boldsymbol{\lambda}
=
(\Phi_1,\Phi_2,\Phi_1^{+},\Phi_2^{+}),
\label{eq:positive-p-expansion}
\end{align}
where
\begin{align}
\hat{\Lambda}(\boldsymbol{\lambda})
&=
\prod_{j=1}^{2}
\frac{
|\Phi_j\rangle
\langle(\Phi_j^{+})^{*}|
}{
\langle(\Phi_j^{+})^{*}|\Phi_j\rangle
}
\label{eq:positive-p-kernel}
\end{align}

is the normalized off-diagonal coherent-state kernel, and $\Phi_j$, $\Phi_j^+$ are independent complex phase-space fields associated with
the two optical channels. A stochastic trajectory denotes one realization of the joint evolution of $(\Phi_1,\Phi_2,\Phi_1^+,\Phi_2^+)$, generated by a particular realization of the stochastic noise fields introduced below. The reservoir variables do not
appear in the phase-space vector because they have already
been traced out; their influence is retained through the
dissipative terms in the reduced field evolution. In
particular, $\Phi_j^{+}$ is not constrained to equal
$\Phi_j^{*}$ on an individual stochastic trajectory.
Normally ordered operator moments are recovered by averaging
the corresponding products of $\Phi_j$ and $\Phi_j^{+}$ over
the stochastic ensemble.

The continuum fields are normalized such that $\Phi_j^{+}\Phi_j$ is a photon flux and $\hbar\omega_j\Phi_j^{+}\Phi_j$ is an optical power. We introduce a retarded frame moving at the group velocity of channel~1,
\begin{equation}
t' = t-\frac{z}{v_1},
\qquad
z'=z,
\label{eq:retarded-frame}
\end{equation}
and define the dimensionless coordinates
\begin{equation}
\tau=\frac{t'}{t_0},
\qquad
\zeta=\frac{z'}{z_0},
\qquad
z_0=\frac{t_0^2}{|k''^{(1)}|},
\label{eq:dimensionless-coordinates}
\end{equation}
where $t_0$ is the characteristic temporal scale and $z_0$ is the corresponding dispersion length of channel~1.

The two positive-$P$ fields in channel $j$ are scaled as
\begin{equation}
\bigl(\Phi_j,\Phi_j^{+}\bigr)(z',t')
=
\sqrt{\frac{P_j}{\hbar\omega_j}}\,
\bigl(u_j,u_j^{+}\bigr)(\zeta,\tau).
\label{eq:field-scaling}
\end{equation}

Here, $P_j$ is a fixed reference power used to normalize channel $j$, rather than a trajectory- or time-dependent quantity. For the input fields introduced in Section~\ref{sec:input-and-metrics}, $P_1$ is chosen as the peak power of the weak quantum pulse, while $P_2$ is the peak launch power of the classical waveform.

The dimensionless walk-off and dispersion parameters are
\begin{equation}
W
=
\tfrac{v_2-v_1}{v_1v_2}
\tfrac{z_0}{t_0},\ \,
s_1=\operatorname{sgn}(k''^{(1)}),\ \,
D_2=\tfrac{k''^{(2)}}{|k''^{(1)}|}
\left(\tfrac{v_2}{v_1}\right)^2.
\label{eq:linear-dimensionless-parameters}
\end{equation}

The first-order temporal-derivative term $W\frac{\partial }{\partial\tau}$ accounts for the walk-off of channel 2 in the retarded frame moving with channel~1.

For closely spaced carriers, $v_2/v_1\simeq1$, and $D_2$ reduces approximately to the ratio of the two dispersion coefficients.

Let $\chi_1$ and $\chi_2$ denote the continuum self-Kerr coefficients obtained from the corresponding transverse-mode overlap integrals. The dimensionless nonlinear coefficients are
\begin{align}
G_{11}
&=
\frac{\chi_1z_0P_1}{\hbar\omega_1},
&
G_{22}
&=
\frac{\chi_2z_0P_2}{\hbar\omega_2},
\label{}
\\
G_{12}
&=
\frac{2\sqrt{\chi_1\chi_2}\,z_0P_2}
{\hbar\omega_2},
&
G_{21}
&=
\frac{2\sqrt{\chi_1\chi_2}\,z_0P_1}
{\hbar\omega_1}.
\label{eq:cross-kerr-coefficients}
\end{align}
The factor of two in the cross-channel coefficients is the conventional XPM enhancement for fields with the same polarization and transverse-mode overlap. For spectrally close channels with a common classical nonlinear coefficient $\gamma_{\mathrm{NL}}$, these expressions reduce approximately to
\begin{equation}
G_{jj}\simeq\gamma_{\mathrm{NL}}P_jz_0,\quad
(G_{12},G_{21})
\simeq2\gamma_{\mathrm{NL}}z_0(P_2,P_1).
\label{eq:classical-nonlinear-limit}
\end{equation}

\subsection{Coupled stochastic propagation equations}
\label{subsec:coupled-propagation-equations}

The positive-$P$ Fokker--Planck equation yields the coupled stochastic partial differential equations in the It\^{o} convention. For numerical propagation, however, we employ a semi-implicit midpoint scheme formulated in the Stratonovich convention. We therefore convert the It\^{o} equations to their equivalent Stratonovich form before numerical integration. The corresponding drift corrections and their discretized expressions are provided in the Supplementary Material.
 For compactness, define the linear propagation operators
\begin{equation}
\begin{aligned}
\mathcal{L}_1
&=
-\overline{\alpha}_{\mathrm f}
-\frac{\mathrm{i}s_1}{2}\frac{\partial^2}{\partial\tau^2},
\\
\mathcal{L}_2
&=
-\overline{\alpha}_{\mathrm f}
+W\frac{\partial}{\partial\tau}
-\frac{\mathrm{i}D_2}{2}\frac{\partial^2}{\partial\tau^2}.
\end{aligned}
\label{eq:linear-operators}
\end{equation}

The dimensionless field-loss coefficient is
$\overline{\alpha}_{\mathrm f}
=\alpha_{\mathrm f}z_0
=[\ln(10)/20]a_{\mathrm{dB}}z_0$, where
$a_{\mathrm{dB}}$ is the power attenuation in
$\mathrm{dB}$ per unit length.

With these definitions, the two positive-$P$
fields obey
\begin{align}
\frac{\partial u_1}{\partial\zeta}
&=
\mathcal{L}_1u_1
+\mathrm{i}G_{11}u_1^2u_1^{+}
+\mathrm{i}G_{12}u_1u_2u_2^{+}
+R_1
\nonumber\\
&\quad+
\sqrt{\frac{\mathrm{i}\chi_1z_0}{t_0}}\,
u_1\eta_1,
\label{eq:u1-propagation}
\\
\frac{\partial u_2}{\partial\zeta}
&=
\mathcal{L}_2u_2
+\mathrm{i}G_{22}u_2^2u_2^{+}
+\mathrm{i}G_{21}u_1u_1^{+}u_2
+R_2
\nonumber\\
&\quad+
\sqrt{\frac{4\mathrm{i}\chi_2z_0}{t_0}}\,
u_2\eta_1
+
\sqrt{\frac{-3\mathrm{i}\chi_2z_0}{t_0}}\,
u_2\eta_2.
\label{eq:u2-propagation}
\end{align}

Here, $R_j$ denotes the dimensionless Langevin field
associated with the Markov loss reservoir of channel $j$.
The equations for $u_1^{+}$ and $u_2^{+}$ follow from the
formal substitutions
$u_j\leftrightarrow u_j^{+}$,
$\mathcal{L}_j\rightarrow\mathcal{L}_j^{+}$,
$\mathrm{i}\rightarrow-\mathrm{i}$,
$\eta_r\rightarrow\eta_r^{+}$, and
$R_j\rightarrow R_j^{+}$.
This relation does not imply
$u_j^{+}=u_j^{*}$ on individual trajectories.
The scaled reservoir fields are
\begin{equation}
R_j
=
\Gamma^{(j)}
\sqrt{\frac{\hbar\omega_jz_0^2}{P_jv_j}},
\qquad j\in\{1,2\},
\label{eq:scaled-reservoir-fields}
\end{equation}

The Kerr noises are independent real Gaussian fields with
zero mean and correlations
\begin{align}
\langle\eta_r\eta_s'\rangle_{\mathrm{st}}
&=
\langle\eta_r^{+}\eta_s^{+\prime}\rangle_{\mathrm{st}}
=
\delta_{rs}\delta(\zeta-\zeta')
\delta(\tau-\tau'),
\nonumber\\
\langle\eta_r\eta_s^{+\prime}\rangle_{\mathrm{st}}
&=0,
\label{eq:kerr-noise-correlations}
\end{align}
where a prime indicates evaluation at $(\zeta',\tau')$.

The scaled reservoir fields also have zero mean. For
independent thermal Markov reservoirs, their nonzero
normally ordered correlation is
\begin{equation}
\left\langle
R_j^{+}(\zeta,\tau)R_k(\zeta',\tau')
\right\rangle_{\mathrm{st}}
=
\delta_{jk}
\frac{2\overline{\alpha}_{\mathrm f}n_{\mathrm{th},j}}
{N_{0j}}
\delta(\zeta-\zeta')\delta(\tau-\tau'),
\label{eq:reservoir-noise-correlation}
\end{equation}
where
$N_{0j}=P_jt_0/(\hbar\omega_j)$ and
$n_{\mathrm{th},j}$ is the thermal reservoir occupation.
The anomalous correlations
$\langle R_jR_k\rangle_{\mathrm{st}}$ and
$\langle R_j^{+}R_k^{+}\rangle_{\mathrm{st}}$ vanish.

The two Kerr-noise sets ${\eta_r}$ and ${\eta_r^{+}}$ are statistically independent. However, the same realization of $\eta_1$ appears in Eqs.~\eqref{eq:u1-propagation} and \eqref{eq:u2-propagation}, generating the off-diagonal diffusion associated with the cross-Kerr interaction. Although the chosen factorization of the diffusion matrix is not unique, all valid factorizations reproduce the same stochastic ensemble moments. The stochastic diffusion terms are a phase-space representation of Kerr-generated quantum correlations; they are distinct from the Langevin fields associated with dissipative reservoirs.

The reservoir Langevin fields $R_j$ and $R_j^{+}$ are independent of the Kerr noises and of the reservoir fields associated with the other channel. The Markov loss reservoirs contribute both the deterministic attenuation contained in $\mathcal{L}_j$ and the fluctuations described by Eq.~\eqref{eq:reservoir-noise-correlation}. At optical carrier frequencies and room temperature, $n_{\mathrm{th},j}
=[\exp(\hbar\omega_j/k_{\mathrm B}T)-1]^{-1}\ll1$, and the normally ordered thermal correlation is therefore strongly suppressed.

\subsection{Physical interpretation and range of validity}
\label{subsec:model-interpretation}

The terms proportional to $G_{11}$ and $G_{22}$ describe SPM within the quantum and classical channels, respectively. The term proportional to $G_{12}$ describes the nonlinear influence of the classical field on channel~1, whereas $G_{21}$ describes the reciprocal influence of the quantum field on channel~2. Because $P_2\gg P_1$ in the regime considered here, $G_{12}$ is much larger than $G_{21}$. The classical channel can therefore impose an appreciable nonlinear phase on the weak quantum signal, while the back-action of the quantum signal on the classical waveform remains negligible.

The retained XPM interaction is number conserving and does not directly transfer optical energy between the two carrier-frequency channels. Instead, the modulated classical intensity produces a time-dependent phase shift on the quantum field. Chromatic dispersion subsequently converts this phase modulation into temporal displacement and deformation of the quantum pulses, while the symbol-dependent classical waveform produces pulse-to-pulse variations in the induced phase. These effects motivate the temporal-mode and interferometric metrics introduced in Section~\ref{sec:input-and-metrics}.

The multiplicative stochastic terms in Eqs.~\eqref{eq:u1-propagation}--\eqref{eq:u2-propagation} represent the quantum diffusion generated by the Kerr interaction. They are not an externally imposed classical noise source. Physical field moments are obtained through ensemble averages; for example,
\begin{align}
&\left\langle
:
[\hat{\Phi}_1^\dagger\hat{\Phi}_1]^m
[\hat{\Phi}_2^\dagger\hat{\Phi}_2]^n
:
\right\rangle
=
C_{1}^mC_{2}^n
\left\langle
(u_1^{+}u_1)^m
(u_2^{+}u_2)^n
\right\rangle_{\mathrm{st}},
\label{eq:normally-ordered-moment}
\end{align}
where 
$C_{j} = {P_j}/{\hbar\omega_j}$
Similarly, the ensemble-averaged complex envelope is proportional to $\langle u_j\rangle_{\mathrm{st}}$, whereas quadrature correlations depend on joint moments of $u_j$ and $u_j^{+}$.

To isolate the effect of the classical channel on channel~1, the reference propagation used throughout this work is set by turning off the classical channel $P_2=0$ while retaining dispersion, walk-off, attenuation, SPM, and the associated positive-$P$ fluctuations. The comparison therefore measures the Kerr-mediated interchannel influence rather than the total deviation from lossless or dispersion-free propagation.

\section{Signal Model and Performance Metrics}
\label{sec:input-and-metrics}

\subsection{Input signal model}
\label{subsec:input-signals}

We consider a weak coherent pulse train in channel~1 and a
data-modulated coherent waveform in channel~2. The dimensionless
coordinates and fields are those introduced in
Eqs.~\eqref{eq:dimensionless-coordinates} and
\eqref{eq:field-scaling}.

\subsubsection{Quantum pulse train}

The input to channel~1 consists of $N_{\mathrm p}$
transform-limited Gaussian pulses,
\begin{equation}
u_1(\tau,0)
=
\sum_{n=1}^{N_{\mathrm p}}
\exp(\mathrm{i}\phi_n)
\exp\left[
-2\ln(2)
\left(
{\tau-\tau_n}
\right)^2
\right].
\label{eq:quantum-input}
\end{equation}
Here,
$\tau_n=\tau_0+(n-1)T_{\mathrm{rep}}/t_0$,
$T_{\mathrm{rep}}=1/f_{\mathrm{rep}}$, and $t_0$ is the intensity FWHM. The mutually coherent
input train has $\phi_n=0$. For a mean photon number $\mu$
per pulse, the normalization power is
$P_1
=\frac{2\sqrt{\ln(2)}}{\sqrt{\pi}}
\frac{\hbar\omega_1\mu}{t_0}$.

\subsubsection{Classical data channel}

Channel~2 carries the \gls{rrc}-shaped $M$-QAM waveform
\begin{equation}
u_2(\tau,0)=s(\tau)
=C_{\mathrm{QAM}}\sum_k c_k h_{\mathrm{RRC}}
\left(\tau-k\widetilde{T}_{\mathrm{s}};\beta\right),
\label{eq:classical-input}
\end{equation}

where ${\tau}_{\mathrm{s}}
=(t_0f_{\mathrm{sym}})^{-1}$ and $c_k$ are the equiprobable points on the symbol constellation and $\beta$ is the \gls{rrc} roll-off factor. The factor $C_{\mathrm{QAM}}$ scales the waveform such that $|s(\tau)|  \lesssim 1$. The simulations are instead parameterized by the average classical launch power, $P_2^{\mathrm{(avg)}}=C_{\mathrm{avg}}P_2$, where $C_{\mathrm{avg}}$ is obtained by averaging $|s(\tau)|^2$ over the simulated time window. Further details of the constellation, pulse shaping, and power normalization are provided in the Supplementary Material.


Conditioned on a particular QAM sequence, both input fields are
multimode coherent states. Their positive-$P$ variables therefore
have the deterministic initial values
\begin{equation}
u_j^+(\tau,0)=u_j^*(\tau,0) \qquad j \in \{1,2\}.
\label{eq:positive-p-initial-conditions}
\end{equation}
After propagation begins, $u_2$ and $u_2^+$ evolve as independent
stochastic variables and are not complex conjugates on individual
trajectories. Carrier-frequency factors do not appear explicitly in
Eqs.~\eqref{eq:quantum-input} and \eqref{eq:classical-input} because
the envelopes are defined in their respective carrier-frequency
rotating frames.

\subsection{WDM multiplexing and channel selection}
\label{subsec:MUX}

The WDM interfaces in Fig.~1 are modeled by Gaussian filters that provide $\Lambda_{\mathrm{WDM}}$ dB of power isolation from the adjacent channel at spacing $\Delta\omega_{\mathrm{WDM}}$:
\begin{equation}
H_j(\omega)=
\exp\left[
-\frac{\ln(10)}{20}\Lambda_{\mathrm{WDM}}
\left(\frac{\omega-\omega_0^{(j)}}{\Delta\omega_{\mathrm{WDM}}}\right)^2
\right],
\end{equation}

where $\Delta\omega_{\mathrm{WDM}}/(2\pi)=100\,\mathrm{GHz}$ and $\Lambda_{\mathrm{WDM}}=30\,\mathrm{dB}$ denote the channel spacing and nominal adjacent-channel power isolation, respectively. Thus, $|H_j(\omega_j\pm\Delta\omega_{\mathrm{WDM}})|^2=10^{-\Lambda_{\mathrm{WDM}}/10}$. The Gaussian filtering operation is applied at the WDM multiplexing and demultiplexing interfaces to represent physical channel preparation and selection. In the adopted channel-resolved representation, however, the two propagated envelopes occupy distinct carrier-frequency subspaces. Consequently, the filtering operation does not introduce coupling between the simulated channels and does not affect the reported observables. Residual demultiplexer leakage into the detected quantum channel is not included; the reported results therefore isolate Kerr-mediated effects.

\subsection{Reference propagation}
\label{subsec:reference-propagation}

At each distance, the co-propagating quantum field is compared with a quantum-only dedicated-fiber reference, termed here the dark-fiber case, in which the same quantum signal propagates without classical traffic. The reference is obtained by setting $P_2=0$, while retaining linear attenuation, chromatic dispersion, SPM, and the associated positive-$P$ fluctuations. In the present model, this case provides the best-case baseline for the mean quantum-field evolution because it excludes Kerr perturbations induced by classical channels. All remaining parameters are unchanged, so deviations from this reference isolate the Kerr-mediated influence of the classical channel.

\subsection{Performance metrics}
\label{subsec:performance-metrics}

Three complementary quantities are used to characterize distinct
aspects of the propagated quantum field: the quadrature-fluctuation
spectrum, the complex temporal-mode overlap, and the adjacent-pulse
interference visibility. They respectively probe changes in quantum
field fluctuations, deformation of the ensemble-averaged complex
envelope, and degradation of pulse-to-pulse coherence.

\subsubsection{Quadrature-fluctuation spectrum}
\label{subsubsec:quadrature-spectrum} Following \cite[Eqn. (7.1)]{drummond1987quantum} we choose a $\operatorname{LO}$ for homodyning the field, subtract the mean of the resulting photocurrent difference and compute the Fourier spectrum to obtain the quadrature-fluctuation spectrum.

For positive-$P$ trajectory $r$, the LO-normalized quadrature and its fluctuation Fourier component are
\begin{align}
X_{\theta}^{(r)}(t,z)
&=
e^{\mathrm{i}\theta}\Phi_1^{+(r)}(t,z)\Phi_{\mathrm{LO}}(t)
+
e^{-\mathrm{i}\theta}\Phi_{\mathrm{LO}}^*(t)\Phi_1^{(r)}(t,z),
\label{eq:homodyne-estimator}\\
\delta X_{\theta}^{(r)}(\omega,z)
&=
\frac{1}{\sqrt{2\pi}}
\int_{\mathcal W}
\left[
X_{\theta}^{(r)}(t,z)
-\langle X_{\theta}(t,z)\rangle_{\mathrm{st}}
\right]
e^{-\mathrm{i}\omega t}\,\mathrm{d}t .
\label{eq:homodyne-fourier}
\end{align}
Here, $\theta$ is the LO phase, $\omega$ is the homodyne analysis frequency, and $\mathcal W$ is the detection window.
To isolate changes in the quantum-field fluctuations from mean-field mode deformation, the LO is modeled as an independently prepared strong coherent field whose temporal mode is matched to the ensemble-averaged propagated signal. Defining
\begin{equation}
\overline{\Phi}_1(t,z)
=
\left\langle \Phi_1(t,z)\right\rangle_{\mathrm{st}},
\end{equation}
the normalized LO mode is
\begin{equation}
f_{\mathrm{LO}}(t,z)
=
\frac{\overline{\Phi}_1(t,z)}
{\left[\displaystyle\int_{\mathcal W}
\left|\overline{\Phi}_1(t,z)\right|^2dt\right]^{1/2}},
\end{equation}
where $\int_{\mathcal W} |f_{\mathrm{LO}}(t,z)|^2dt=1.$

For each propagation distance and classical launch power, $f_{\mathrm{LO}}$ is calculated from the ensemble mean and subsequently held fixed for all stochastic trajectories; the phase $\theta$ is scanned independently to identify the extremal quadratures. Thus, the LO follows the mean complex envelope but not its trajectory-dependent quantum fluctuations.

We assume a strong pulsed LO whose normalized temporal mode $f_{\mathrm{LO}}(t)$ is matched to the stochastic ensemble-averaged signal field. In the strong-LO limit, the normalized quadrature-fluctuation spectrum is independent of the LO amplitude; hence, only its normalized temporal mode enters the stochastic estimator. We therefore set $\Phi_{\mathrm{LO}}(t)=f_{\mathrm{LO}}(t)$ when evaluating the normalized spectrum.

The normally ordered fluctuation spectrum and the variance are
\begin{align}
S_{\theta}^{P}(\omega,z)
&=
2\pi
\left\langle
\delta X_{\theta}(-\omega,z)
\delta X_{\theta}(\omega,z)
\right\rangle_{\mathrm{st}},
\label{eq:positive-p-quadrature-spectrum}\\
V_{\theta}(\omega,z)
&=
1+S_{\theta}^{P}(\omega,z).
\label{eq:quadrature-spectrum-snu}
\end{align}
The variance spectrum $V_{\theta} \geq 0$ is expressed in shot-noise units (SNU), with the shot-noise level normalized to $V_{\theta}=1$. Hence, $S_{\theta}^{P}=V_{\theta}-1$ is the signed deviation from shot noise: $S_{\theta}^{P}=0$ corresponds to the shot-noise level, $S_{\theta}^{P}<0$ denotes sub-shot-noise fluctuations (quadrature squeezing), and $S_{\theta}^{P}>0$ denotes excess fluctuations above shot noise \cite{drummond1987quantum}. Because the doubled positive-$P$ fields are independent, $\delta X_{\theta}(-\omega,z)$ is evaluated directly from the negative-frequency bin rather than replaced by $\delta X_{\theta}^{*}(\omega,z)$.

The LO-phase scan selects $\theta_{\min}$ and $\theta_{\max}$, which yield the minimum and maximum quadrature spectra, respectively. Holding each selected phase fixed over the full frequency range, we define
\begin{align}
S_{\min}^{P}(\omega,z)
= S_{\theta_{\min}}^{P}(\omega,z),
\ \
S_{\max}^{P}(\omega,z)
= S_{\theta_{\max}}^{P}(\omega,z).
\label{eq:maximum-quadrature-spectrum}
\end{align}

\subsubsection{Complex temporal-mode overlap}
\label{subsubsec:temporal-mode-overlap}
The complex temporal-mode overlap measures preservation of each pulse's complex temporal envelope relative to dark-fiber propagation; a reduction indicates temporal amplitude or phase deformation that would produce mode mismatch in a receiver configured for the reference pulse.

Within the fixed window $\mathcal{W}_n$ assigned to pulse $n$, let
$f_n(\tau,z)=\langle u_1(\tau,z)\rangle_{\mathrm{st}}$ and
$f_{n,\mathrm{ref}}(\tau,z)$ denote the ensemble-averaged
co-propagating and reference envelopes {($P_2=0$)}, respectively. Their normalized
temporal-mode overlap is
\begin{equation}
\mathcal{F}_n(z)
=
\tfrac{
\left|
\int_{\mathcal{W}_n}
f_{n,\mathrm{ref}}^*(\tau,z)f_n(\tau,z)\,\mathrm{d}\tau
\right|^2
}{
\left[
\int_{\mathcal{W}_n}
|f_{n,\mathrm{ref}}(\tau,z)|^2\,\mathrm{d}\tau
\right]
\left[
\int_{\mathcal{W}_n}
|f_n(\tau,z)|^2\,\mathrm{d}\tau
\right]
}.
\label{eq:temporal-mode-overlap}
\end{equation}
By the Cauchy--Schwarz inequality, $0\leq\mathcal{F}_n\leq1$, with
$\mathcal{F}_n=1$ when the two envelopes differ only by a constant
complex factor. Thus, the normalization removes uniform attenuation
and global phase rotation while retaining sensitivity to temporal
displacement, broadening, amplitude deformation, and time-dependent
phase modulation.

For a pulse train consisting of $N_p$ pulses, the pulse-averaged overlap is $\overline{\mathcal{F}}(z)$ and its pulse-to-pulse sample standard deviation is denoted by $s_{\mathcal{F}}(z)$.

The results use $1-\overline{\mathcal{F}}$ as the temporal-mode
penalty; the $s_{\mathcal{F}}$ band quantifies symbol-dependent
pulse-to-pulse variation, not numerical or stochastic uncertainty in
the ensemble mean.

\subsubsection{Adjacent-pulse interference visibility}
\label{subsubsec:interference-visibility}

The adjacent-pulse interference visibility \cite{kobayashi2014evaluation} quantifies how well relative phase coherence is preserved between successive quantum pulses, as measured by the fringe contrast obtained when one pulse is delayed by $T_{\mathrm{rep}}$ and interfered with the next. After aligning pulses $n$ and $n+1$ in the local time $\vartheta\in\mathcal{W}_0$, their photon numbers and first-order mutual coherence are
\begin{align}
N_n(z)
&=
N_{01}\int_{\mathcal{W}_0}
\left\langle
u_{1,n}^+(\vartheta,z)u_{1,n}(\vartheta,z)
\right\rangle_{\mathrm{st}}
\,\mathrm{d}\vartheta,
\label{eq:aligned-pulse-photon-number}\\
G_{n,n+1}^{(1)}(z)
&=
N_{01}\int_{\mathcal{W}_0}
\left\langle
u_{1,n}^+(\vartheta,z)u_{1,n+1}(\vartheta,z)
\right\rangle_{\mathrm{st}}
\,\mathrm{d}\vartheta.
\label{eq:first-order-mutual-coherence}
\end{align}
Scanning the interferometer phase $\varphi$ gives output intensity values
\begin{align}
I_n(\varphi,z)
&=
N_n(z)+N_{n+1}(z)
+2\operatorname{Re}\left[
e^{\mathrm{i}\varphi}G_{n,n+1}^{(1)}(z)
\right],
\label{eq:interferometer-output}\end{align}
Upon setting $I_{n,\max}(z):=\max_{\varphi}I_{n}(\varphi,z)$ and $I_{n,\min}(z):=\min_{\varphi}I_{n}(\varphi,z)$ we 
arrive at
\begin{align}
\mathcal{V}_n(z) 
&=
\frac{I_{n,\max}(z)-I_{n,\min}(z)}
{I_{n,\max}(z)+I_{n,\min}(z)}
=\frac{2|G_{n,n+1}^{(1)}(z)|}
{N_n(z)+N_{n+1}(z)}.
\label{eq:adjacent-pulse-visibility}
\end{align}
Unity visibility requires perfect first-order coherence, identical temporal modes, and equal pulse energies; differential phase fluctuations, energy imbalance, or mode mismatch reduce $\mathcal{V}_n$.

For $N_{\mathrm p}>2$, the mean visibility is $\overline{\mathcal{V}}(z)$ and pair-to-pair sample standard deviation is $s_{\mathcal{V}}(z)$.

The results report $1-\overline{\mathcal{V}}$ as the visibility penalty, while $s_{\mathcal{V}}$ quantifies symbol-dependent pair-to-pair variation. 
Under ideal balanced-interferometer conditions, the visibility-related error contribution \cite{branciard2005security} solely due to the modeled Kerr-mediated channel co-propagation is
\begin{equation}
e_{\mathrm{vis}}(z)
=\frac{1-\overline{\mathcal{V}}(z)}{2}.
\label{eq:visibility-error-probability}
\end{equation}

\subsection{Numerical evaluation}
\label{subsec:numerical-evaluation}

The four coupled fields $u_1$, $u_1^+$, $u_2$, and $u_2^+$ are integrated using a semi-implicit midpoint method \cite{werner1997robust,kiesewetter2023xspde3}, with dispersion and walk-off evaluated in the frequency domain and the Kerr drift and diffusion terms in the time domain. A $2^{12}$-point temporal grid with $\Delta t=2.4969\,\mathrm{ps}$ gives a $10.2273\,\mathrm{ns}$ window containing 511 QAM symbols and 15 quantum pulses. This spacing corresponds to an effective symbol rate of $49.964\,\mathrm{Gbaud}$, quoted as the nominal $50\,\mathrm{Gbaud}$ rate, and supports periodic continuation of the QAM realization. Circular boundary smoothing prevents discontinuities as walk-off samples the repeated symbol pattern. Propagation is performed with a $10\,\mathrm{m}$ longitudinal step.

For each launch power, all four fields are propagated over $2^{14}$ independent positive-$P$ trajectories. {Stochastic convergence is verified by increasing the ensemble size; the real quadrature spectra stabilize while their residual imaginary components decrease toward zero (Supplementary Material, Sec. S7).} Quantum observables are obtained from trajectory averages, whereas subsequent pulse- or pair-wise statistics quantify QAM-induced sequence variation. Owing to the computational demands of the doubled phase space and large stochastic ensemble, the trajectories are evaluated in parallel on GPUs. The remaining simulation parameters are listed in Table~\ref{tab:simulation_parameters}.

\section{Results}
\label{sec:results}

\begin{table}[!t]
\caption{Simulation Parameters}
\label{tab:simulation_parameters}
\centering
\footnotesize
\setlength{\tabcolsep}{3pt}
\renewcommand{\arraystretch}{1.15}

\begin{threeparttable}
\begin{tabularx}{\columnwidth}{@{}
    c
    >{\raggedright\arraybackslash}X
    >{\centering\arraybackslash}p{0.20\columnwidth}
    >{\centering\arraybackslash}p{0.30\columnwidth}
    @{}}
\toprule
\textbf{No.} &
\textbf{Parameter} &
\textbf{Symbol} &
\textbf{Value} \\
\midrule

1  & Pulse duration
   & $t_0$
   & $100\,\mathrm{ps}$ \\

2  & Mean photons per pulse
   & $\mu$
   & $0.4$ \\

3  & Pulse repetition rate
   & $f_{\mathrm{rep}}$
   & $1.5\,\mathrm{GHz}$ \\

4  & Pulses in window
   & $N_p$
   & $14$ \\

5  & Quantum carrier
   & $\omega_0^{(1)}$
   &  $2\pi \times 193.4\,\mathrm{THz}$ \\

6  & Classical carrier
   & $\omega_0^{(2)}$
   &  $2\pi \times 193.3\,\mathrm{THz}$ \\

7  & Nominal \gls{wdm} isolation
   & $\Delta \omega_{\mathrm{WDM}}$
   &  $2\pi \times 100\,\mathrm{GHz}$ \\

8  & Filter isolation
   & $\Lambda_{\mathrm{WDM}}$
   & $30\,\mathrm{dB}$ \\

9  & QAM order
   & $M$
   & $64$ \\

10 & QAM symbol rate
   & $f_{\mathrm{sym}}$
   & $\simeq 50\, \mathrm{Gbaud}$ \\

11 & RRC roll-off factor
   & $\beta$
   & $0.2$ \\

12  & Fiber length
   & $L$
   & $100\,\mathrm{km}$ \\

13  & Dispersion parameter
   & $-$
   & $17\, \mathrm{ps/(km\,nm)}$ \\

14  & Kerr nonlinear coefficient
   & $\gamma_{\mathrm{NL}}$
   & $0.78\,\mathrm{W}^{-1}\mathrm{km}^{-1}$ \\

15  & Power attenuation
   & $a_{\mathrm{dB}}$
   & $0.2 \, \mathrm{dB/km}$ \\

16  & Temperature
   & $T$
   & $300\,\mathrm{K}$ \\

\bottomrule
\end{tabularx}

\end{threeparttable}
\end{table}



The parameters used in the numerical analysis are summarized in
Table~\ref{tab:simulation_parameters}. Unless stated otherwise, the
quantum channel contains pulses with mean photon number $\mu=0.4$,
temporal width $t_0=100~\mathrm{ps}$, and repetition rate
$f_{\mathrm{rep}}=1.5~\mathrm{GHz}$. The two channels are propagated
over $100~\mathrm{km}$ of fiber, and the classical launch power
$P_2^{\mathrm{(avg)}}$ is varied to determine how the Kerr-mediated
interchannel interaction affects the three observables introduced in
Section~\ref{subsec:performance-metrics}. The zero-power result and the
calculation with {$P_2^{\mathrm{(avg)}}=0$} provide the corresponding reference
propagation for channel~1.

\begin{figure*}[h!]
    \centering
    \includegraphics[width=1.95\columnwidth]
    {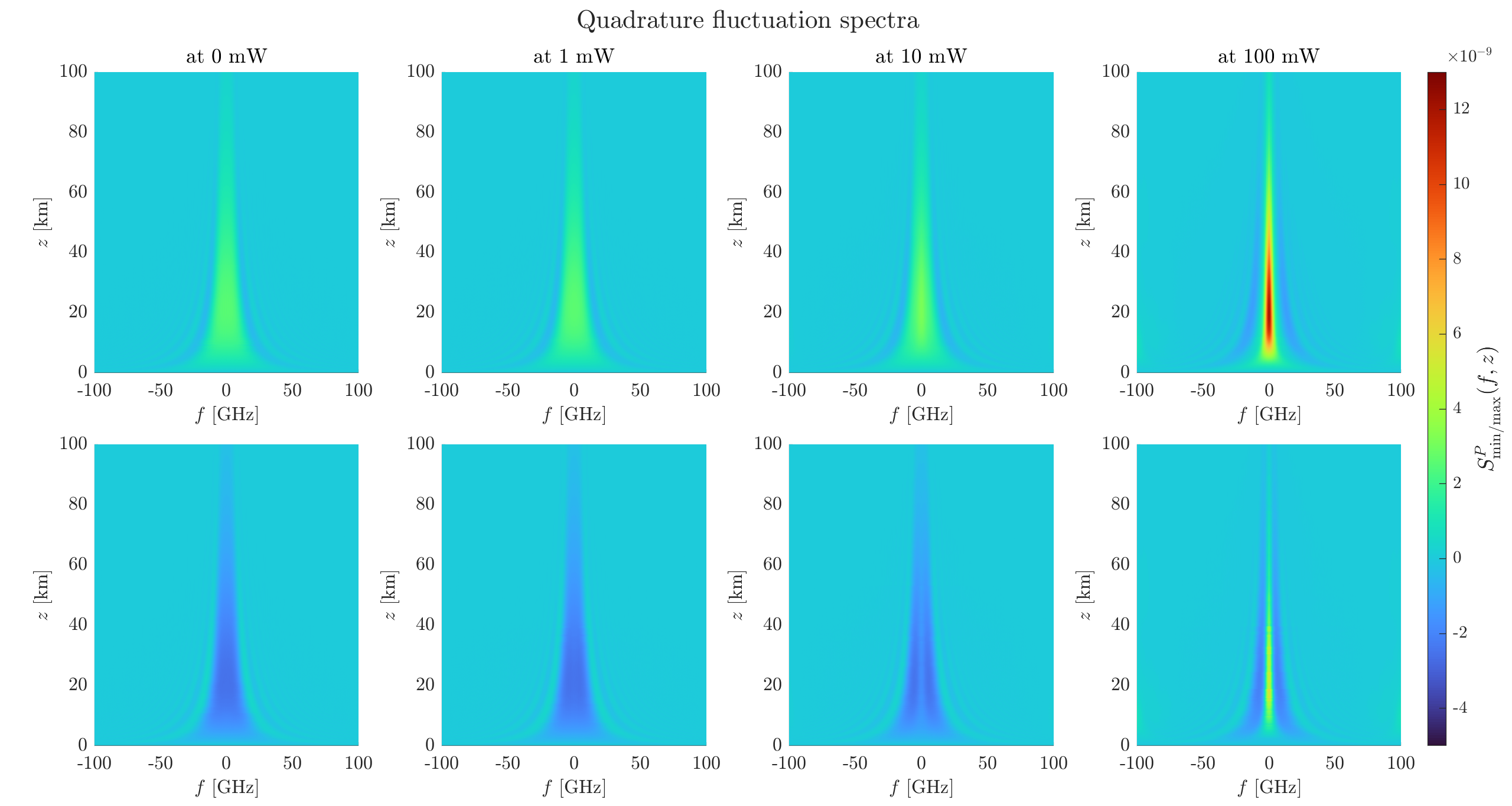}
    \caption{Extremal normally ordered quadrature-fluctuation spectra of the quantum channel as functions of analysis frequency $\omega=2 \pi f$ and propagation distance $z$. The top and bottom rows show $S_{\max}^{P}(\omega,z)$ and $S_{\min}^{P}(\omega,z)$, respectively. From left to right, the columns correspond to classical launch powers of $P_2^{\mathrm{(avg)}}=0, 1, 10\, \text{and}\, 100~\mathrm{mW}$. The spectra are normalized relative to the shot-noise level, for which $S^{P}=0$.
}
    \label{result_M1}
\end{figure*}

\subsection{Quadrature-fluctuation spectrum}
\label{subsec:results-quadrature}
Figure~\ref{result_M1} shows the extremal normally ordered quadrature spectra versus propagation distance and homodyne Fourier analysis frequency $\omega$. The upper and lower rows show $S_{\theta}^{P}$ evaluated at the LO phases selected for its maximum and minimum values, denoted $S_{\max}^{P}$ and $S_{\min}^{P}$, respectively. Each selected phase is used over the full frequency range.

In the absence of the classical channel, the nonlinear evolution of
the weak coherent pulse train produces a positive extremum in
$S_{\max}^{P}$ and a negative extremum in $S_{\min}^{P}$ centered
around zero analysis frequency. These correlations are concentrated
within the bandwidth occupied by the quantum pulse train and evolve
with distance because of the combined action of SPM, dispersion, loss,
and positive-$P$ diffusion. This zero-power reference result establishes the
intrinsic fluctuation spectrum against which the co-propagating cases
are compared.

At $P_2^{\mathrm{(avg)}}=1~\mathrm{mW}$, both extrema remain nearly
indistinguishable from the zero-power result over the full propagation
distance. Increasing the launch power to $10~\mathrm{mW}$ produces a
visible redistribution of the correlations around the central
frequency, but does not create a broadband increase in the
quadrature-noise level. The largest change occurs for
$P_2^{\mathrm{(avg)}}=100~\mathrm{mW}$, where the positive extremum becomes
more strongly localized in frequency and distance. The negative
extremum is also reshaped, although its magnitude changes more weakly
than that of the positive branch.

Importantly, the normally ordered spectra remain on the order of
$10^{-9}$--$10^{-8}$ relative to the unit shot-noise level, even for
the $100~\mathrm{mW}$ stress-test case. Hence,
$V_{\min}=1+S_{\min}^{P}$ and
$V_{\max}=1+S_{\max}^{P}$ remain very close to unity. Classical power
therefore modifies the frequency distribution and the optimal
quadrature angle of the Kerr-generated correlations more strongly than
it changes their absolute shot-noise-normalized magnitude. In
particular, the quadrature response does not scale in direct proportion
to the two-order-of-magnitude increase in $P_2^{\mathrm{(avg)}}$.

The strongest spectral modification develops during the initial part
of the link and subsequently weakens as the fields propagate. This
behavior is consistent with the reduction of the classical power by
fiber attenuation and with dispersive broadening, both of which reduce
the instantaneous nonlinear interaction at larger distances. The
quadrature spectra thus retain information about where along the link
the Kerr interaction is strongest, while remaining comparatively
robust in absolute magnitude.

\begin{figure}
    \centering
    \includegraphics[width=1.0\linewidth]{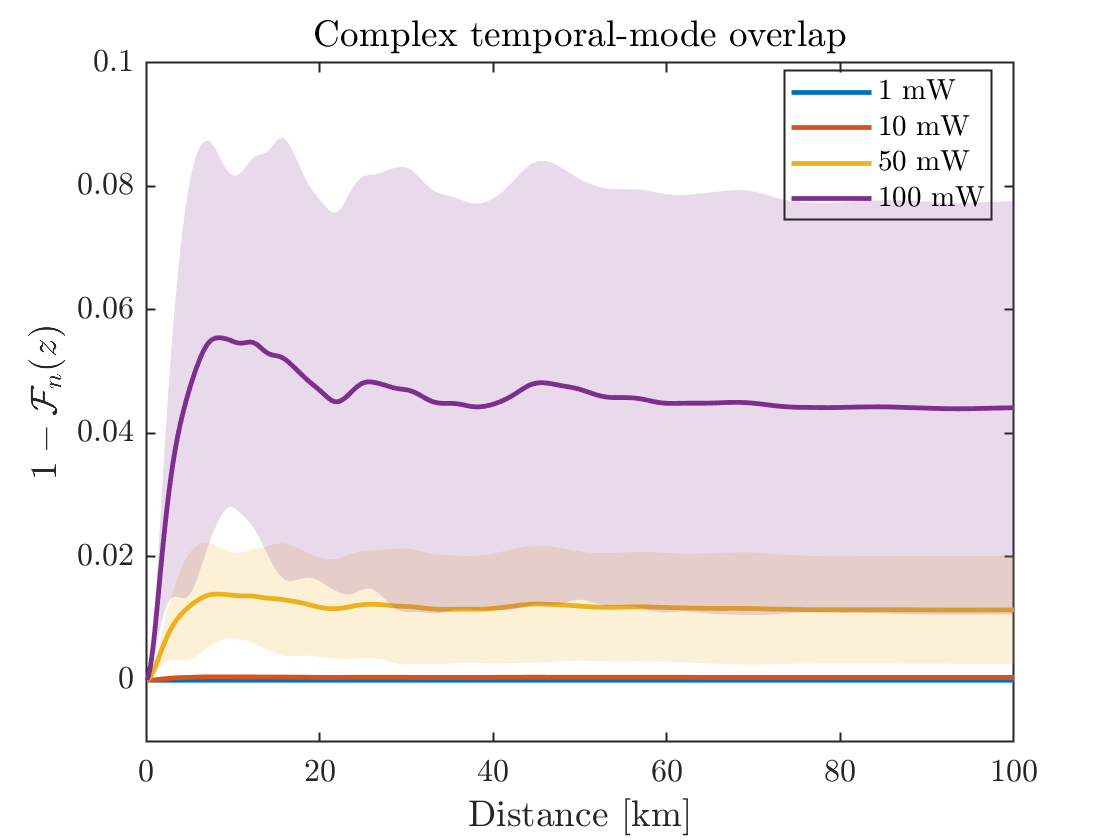}
    \caption{Complex temporal-mode overlap vs distance for the power of the classical channel $(P_2^{\mathrm{(avg)}})$ is 1, 10, 50 and 100 mW. The solid curves show is $1-\overline{\mathcal{F}}(z)$; shaded regions show the pulse-to-pulse sample standard deviation $s_{\mathcal{F}}(z)$.}
    \label{result_M2}
\end{figure}
\subsection{Complex temporal-mode deformation}
\label{subsec:results-temporal-overlap}

The quadrature spectrum does not directly quantify deformation of the
ensemble-averaged pulse envelope. This effect is examined in
Fig.~\ref{result_M2}, which displays the temporal-mode penalty
$1-\overline{\mathcal{F}}(z)$. A value of zero indicates that, within
each pulse window, the co-propagating and reference envelopes differ
only by a constant complex factor. The solid curves show the average
over all pulses, whereas the shaded regions represent the
pulse-to-pulse variation $s_{\mathcal{F}}$ generated by the different
QAM symbols encountered across the simulated sequence.

For average classical powers of $1$ and $10~\mathrm{mW}$, the temporal-mode
penalty remains close to zero throughout the $100~\mathrm{km}$ link.
Within the resolution of the plotted results, the quantum envelope is
therefore preserved at these powers despite the presence of the
classical waveform. A clear departure from the reference mode is
observed at $50~\mathrm{mW}$. The mean penalty increases rapidly over
the first several kilometers, undergoes weak oscillatory evolution,
and approaches approximately $1.2\times10^{-2}$ at
$100~\mathrm{km}$. Thus, the final mean temporal-mode overlap remains
close to $0.988$, but the classical channel produces a resolvable
amplitude- and phase-sensitive deformation.

At $P_2^{\mathrm{(avg)}}=100~\mathrm{mW}$, the effect becomes substantially
larger. The temporal-mode penalty initially rises above
$5\times10^{-2}$, decreases near $25~\mathrm{km}$, and then settles to
approximately $4.5\times10^{-2}$ at the end of the link. The
corresponding mean overlap is therefore about $0.955$. The
nonmonotonic distance dependence results from the accumulated XPM phase
being continually converted into temporal deformation by dispersion,
while attenuation and walk-off change the local interaction strength.
After approximately $50$--$60~\mathrm{km}$, the average penalty varies
only weakly, indicating that most of the observable mode deformation
has accumulated in the earlier part of the fiber.

The shaded regions broaden markedly with increasing launch power. This
shows that the effect is not a uniform deterministic deformation of all
quantum pulses. Instead, the intensity pattern of the QAM waveform
imposes different nonlinear phases on different members of the pulse
train. The resulting symbol-dependent variation is especially
pronounced at $100~\mathrm{mW}$, where some pulse windows experience
considerably more deformation than indicated by the sequence-averaged
overlap. These bands describe physical pulse-to-pulse variation and
should not be interpreted as confidence intervals for the stochastic
ensemble average.

\begin{figure}[ht]
    \centering
    \includegraphics[width=1.0\linewidth]
    {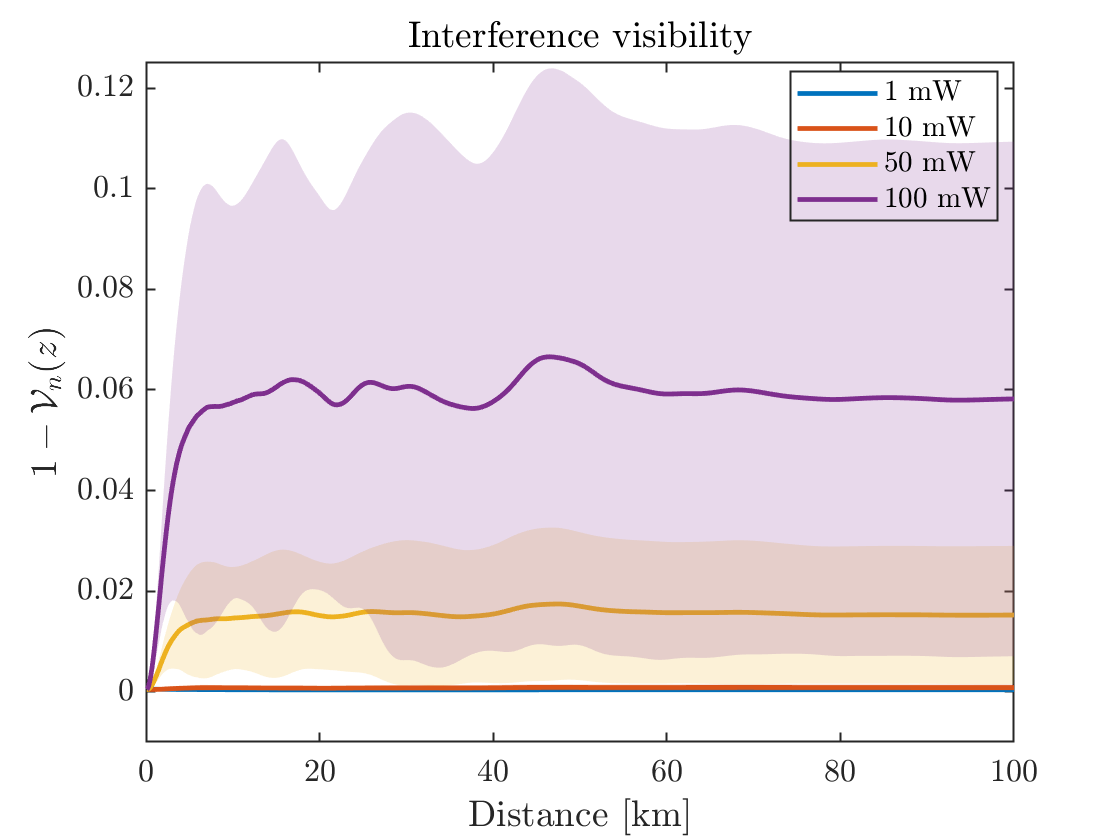}
    \caption{adjacent-pulse visibility vs distance for average classical launch powers $(P_2^{\mathrm{(avg)}})$ of 1, 10, 50 and 100 mW. Solid curve shows $1-\overline{\mathcal{V}}(z)$; shaded regions show the pair-to-pair sample standard deviation $s_{\mathcal{V}}(z)$.}
    \label{result_M3}
\end{figure}
\subsection{Adjacent-pulse interference visibility}
\label{subsec:results-visibility}

Figure~\ref{result_M3} shows the corresponding visibility penalty
$1-\overline{\mathcal{V}}(z)$. This observable is sensitive not only to
the temporal-mode mismatch identified above, but also to differential
phase shifts and unequal energies between successive pulses. It
therefore provides a direct measure of the impairment that would be
observed by a one-pulse-delay interferometer. Such delay interferometers are used for phase-sensitive detection in differential-phase-shift QKD \cite{diamanti2006100} and phase-encoded BB84 \cite{li2020improving}.


The visibility penalty remains negligible for average classical launch powers
of $1$ and $10~\mathrm{mW}$, in agreement with the temporal-mode
overlap. At $50~\mathrm{mW}$, it rises to approximately
$1.5\times10^{-2}$ after $100~\mathrm{km}$, corresponding to an
average visibility of about $0.985$. Under the idealized relation in
Eq.~\eqref{eq:visibility-error-probability}, this value gives a
visibility-related error contribution of approximately
$e_{\mathrm{vis}}\simeq7.5\times10^{-3}$.

For $P_2^{\mathrm{(avg)}}=100~\mathrm{mW}$, the visibility penalty rises
rapidly over the first $10~\mathrm{km}$ and then exhibits a weaker
distance-dependent modulation. It reaches a broad maximum of
approximately $6.5\times10^{-2}$ near the middle of the link and
approaches $6.0\times10^{-2}$ at $100~\mathrm{km}$. The final average
visibility is therefore approximately $0.94$, which corresponds to
$e_{\mathrm{vis}}\simeq3.0\times10^{-2}$ under ideal interferometer
conditions. As in the temporal-mode result, the pair-to-pair variation
increases strongly with power, demonstrating that the QAM-induced phase
perturbation is not identical for every adjacent pulse pair.

The visibility penalty is consistently larger than the temporal-mode penalty
at the two highest powers. At $100~\mathrm{mW}$, for example,
the final penalties are approximately $6.0 \%$ and $4.5\%$,
respectively. This difference is expected because the normalized
temporal overlap is insensitive to a constant phase rotation of an
individual pulse, whereas adjacent-pulse interference responds to the
relative phase between successive pulses. The visibility therefore
captures protocol-relevant phase perturbations that are not fully
represented by the single-pulse temporal-mode overlap.

\subsection{Power threshold for mode and visibility degradation}
\label{subsec:power-threshold}

\begin{figure}[!t]
    \centering
    \includegraphics[width=\columnwidth]{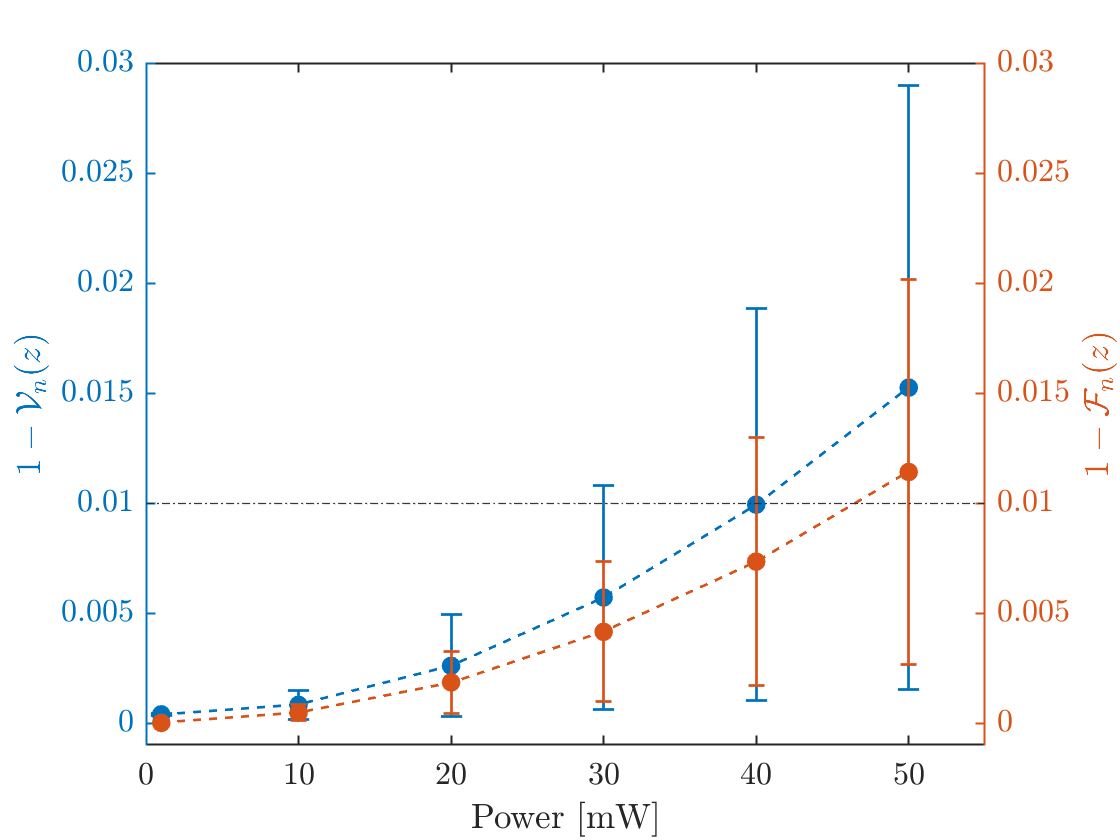}
    \caption{Final-link visibility penalty $1-\overline{\mathcal V}(L)$ and
    temporal-mode penalty $1-\overline{\mathcal F}(L)$ as functions of the
    classical launch power $P_2^{\mathrm{(avg)}}$ for $L=100\,\mathrm{km}$. The markers denote the
    averages over the simulated pulse sequence, and the vertical bars show pulse-to-pulse standard deviations for $\overline{\mathcal F}$ and pair-to-pair standard deviations for $\overline{\mathcal V}$. They do not represent
    confidence intervals or numerical uncertainty. The dash-dotted line marks
    the $1\%$ degradation criterion, while the dashed curves indicate trends between the simulated points.}
    \label{fig:power-threshold}
\end{figure}

To resolve the onset of appreciable degradation more precisely, additional
simulations were performed at intermediate classical launch powers. We define
the final-link visibility and temporal-mode penalties as
\begin{align}
    \epsilon_{\mathcal V}(P_2^{\mathrm{(avg)}})
    &=1-\overline{\mathcal V}(L;P_2^{\mathrm{(avg)}}),\\
    \epsilon_{\mathcal F}(P_2^{\mathrm{(avg)}})
    &=1-\overline{\mathcal F}(L;P_2^{\mathrm{(avg)}}),
    \label{eq:final-link-penalties}
\end{align}
where $L=100\,\mathrm{km}$. Figure~\ref{fig:power-threshold} shows that both
penalties grow nonlinearly with $P_2^{\mathrm{(avg)}}$. Their mean values remain below
$0.6\%$ through $30\,\mathrm{mW}$, increase to approximately $1.0\%$ and
$0.7\%$, respectively, at $40\,\mathrm{mW}$, and reach approximately $1.52\%$
and $1.14\%$ at $50\,\mathrm{mW}$. The visibility is therefore slightly more
sensitive to the classical power than the temporal-mode overlap, consistent
with its additional sensitivity to pulse-to-pulse phase and energy imbalance.

For a metric $M\in\{\mathcal V,\mathcal F\}$, a system-design threshold can be
defined by
\begin{equation}
    P_{1\%}^{(M)}
    =
    \inf\left\{
        P_2^{\mathrm{(avg)}}:1-\overline{M}(L;P_2^{\mathrm{(avg)}})\geq0.01
    \right\}.
    \label{eq:one-percent-threshold}
\end{equation}
The sampled data give $P_{1\%}^{(\mathcal V)}\simeq40\,\mathrm{mW}$. A
piecewise-linear interpolation between the $40$ and $50\,\mathrm{mW}$ points
gives $P_{1\%}^{(\mathcal F)}\simeq46\,\mathrm{mW}$. Requiring both mean
penalties to remain below $1\%$ therefore yields the joint design threshold
\begin{equation}
    P_{1\%}^{(\mathrm{joint})}
    =
    \min\left\{
        P_{1\%}^{(\mathcal V)},P_{1\%}^{(\mathcal F)}
    \right\}
    \simeq40\,\mathrm{mW}.
    \label{eq:joint-one-percent-threshold}
\end{equation}
For the visibility metric, a $1\%$ penalty corresponds to an ideal
visibility-related error contribution of $e_{\mathrm{vis}}=0.5\%$. Because
the error bars quantify pulse-to-pulse variation, rather than uncertainty in
the mean, a design that imposes a per-pulse requirement should include
additional operating margin below the mean-based threshold in
Eq.~\eqref{eq:joint-one-percent-threshold}.

\section{Discussion}
\label{sec:discussion}

\subsection{Origin of the metric-dependent response}
\label{subsec:discussion-physical-origin}

The central finding is a metric-dependent response to classical co-propagation. Classical-field coexistence and crosstalk treatments commonly quantify impairments through noise photons coupled into or generated within the quantum channel \cite{choi2010quantum,patel2012coexistence,peters2009dense,eraerds2010quantum,kumar2015coexistence, qi2010feasibility, du2020impact}. Our model exposes a distinct, photon-number-preserving impairment pathway: Kerr-mediated XPM transfers no photons between the channel subspaces, yet modifies the phase and temporal mode of the quantum field, thereby degrading its coherence and modal fidelity. Although classical propagation equations can describe the corresponding mean-field phase modulation and temporal reshaping, the quantum-stochastic treatment additionally provides the normally ordered correlations required for shot-noise-referenced quadrature-fluctuation spectra. These quantum-noise correlations cannot be obtained from classical field equations alone.
The positive-$P$ framework provides both mean-field deformation and quantum-fluctuation observables within one model, constituting a central novelty of this work. At elevated classical powers, the temporal-mode and visibility penalties reach percent level, whereas the extremal quadrature spectra remain close to shot noise.

This separation arises because the number-conserving Kerr interaction primarily imposes a QAM-dependent temporal phase on the quantum field without transferring photons between carrier channels. Dispersion converts this phase modulation into pulse deformation, while walk-off determines the portion of the classical waveform sampled by each quantum pulse. Consequently, the temporal-mode overlap is reduced by envelope deformation and time-dependent phase, whereas visibility is additionally sensitive to relative phase and energy imbalance between adjacent pulses, explaining its slightly lower $1\%$ threshold. By contrast, the quadrature spectrum probes normally ordered fluctuations about the mean field; increasing $P^{\mathrm{(avg)}}_2$ modifies their spectral distribution and the LO phases associated with the extrema more strongly than their shot-noise-referenced magnitude.

The nonmonotonic distance dependence results from competition among XPM, dispersion, walk-off, and attenuation. XPM is strongest near the fiber input, while loss progressively weakens the interaction and dispersion continues converting the accumulated phase into temporal deformation. The penalties therefore grow rapidly initially, exhibit moderate oscillations, and evolve more slowly at longer distances.

\subsection{Implications for quantum--classical coexistence}
\label{subsec:discussion-implications}

The extremal quadrature noise spectra exhibit pronounced resilience to Kerr-mediated crosstalk, remaining effectively at the unit shot-noise level even at $P_2^{\mathrm{(avg)}}=100~\mathrm{mW}$. This suggests that, for the modeled standard-fiber link, the quadrature-noise magnitude is preserved over launch powers relevant to classical co-propagation. This robustness contrasts with the percent-level temporal-mode and visibility penalties, indicating that mode deformation and coherence loss occur well before appreciable growth in quadrature noise. In Kerr-mediated crosstalk, the quadrature-noise magnitude alone may therefore be insufficient for assessing phase-sensitive or mode-selective quantum links. 

For the investigated $100~\mathrm{km}$ link, the $1\%$ mean-penalty thresholds are approximately $40~\mathrm{mW}$ for visibility and $46~\mathrm{mW}$ for temporal-mode overlap, yielding a joint design threshold near $40~\mathrm{mW}$. This value provides network designers with a quantitative launch-power guideline for quantum--classical co-propagation; operation below it keeps both mean penalties below $1\%$ for the parameters in Table~\ref{tab:simulation_parameters}. Because pulse-to-pulse variation increases near the threshold, practical designs should include an operating margin based on the permitted outage or worst-pulse penalty. Although the numerical threshold is configuration specific, Eqs.~\eqref{eq:one-percent-threshold} and \eqref{eq:joint-one-percent-threshold} provide a transferable procedure for determining coexistence limits in other network configurations.

\subsection{Framework scope and extensibility}
\label{subsec:discussion-limitations}

The objective of this work is to quantify the Kerr-induced noise and field deformation accumulated during quantum-classical co-propagation. The quadrature spectrum, temporal-mode overlap, and adjacent-pulse visibility therefore provide field-level measures that can subsequently be incorporated into protocol-specific QBER or key-rate analyses.

The positive-$P$ formulation provides a unified quantum-mechanical propagation model in which both the weak quantum pulses and strong classical waveform are quantized. Unlike conventional classical or semiclassical treatments, it yields the mean field, first-order coherence, and normally ordered quadrature correlations from the same stochastic evolution, thereby distinguishing mode deformation from quantum-noise generation. The framework can be systematically extended to include delayed Raman interactions and their associated fluctuations, as well as additional channel fields and nonlinear couplings for multichannel WDM systems. The present two-channel Kerr analysis thus provides a modular foundation for broader quantum-classical coexistence studies.

\section{Conclusion}
\label{sec:conclusion}

We developed a two-channel positive-$P$ framework that treats weak quantum pulses and a data-modulated classical signal within a unified quantum-stochastic propagation model. Unlike deterministic classical propagation alone, this approach provides normally ordered quadrature correlations, temporal-mode deformation, and first-order coherence.

For the investigated $100~\mathrm{km}$ link, Kerr-induced quadrature fluctuations remain effectively at the unit shot-noise level up to $100~\mathrm{mW}$, whereas the temporal-mode and visibility penalties reach approximately $4.5\%$ and $6.0\%$, respectively. Their $1\%$ thresholds occur near $46~\mathrm{mW}$ and $40~\mathrm{mW}$, yielding a joint coexistence threshold of approximately $40~\mathrm{mW}$. These results show that Kerr crosstalk degrades temporal mode and pulse-to-pulse coherence well before producing appreciable growth in quadrature noise. Although configuration-specific, the proposed metric-based threshold provides a transferable design criterion for quantum-classical co-propagation networks.

\section*{Acknowledgments}
The authors thank the Q-net-Q Project which has received funding from the European Union’s Digital Europe Programme under grant agreement No 101091732, and is co-funded by the German Federal Ministry of Education and Research (BMBF). This research is part of the Quant-GPlCz project, which is funded by the European Union under ID 101249338  and by the BMDS under grant number 19OS25001F This work was further financed by the DFG via grant NO 1129/2-1 and by the BMBF via grants 16KISQ168, 16KISQ077, 16KISR026 and 16KISQ039.

\IfFileExists{ref.bib}{%
\bibliographystyle{IEEEtran}
\bibliography{ref}
}{}

\newpage

\end{document}


\maketitle

This Supplementary Material provides the detailed derivations and signal definitions supporting the main manuscript. First, we derive the coupled quantum-stochastic propagation model for two co-propagating optical fields in a Kerr fiber, including dispersion, group-velocity mismatch, attenuation, self- and cross-phase modulation, and the associated positive-$P$ diffusion. Second, we formulate the balanced-homodyne quadrature-fluctuation spectrum and its shot-noise normalization for evaluation from the propagated positive-$P$ fields. Finally, we describe the generation and normalization of the root-raised-cosine-shaped $M$-QAM waveform used as the classical-channel input. Together, these sections provide the theoretical and numerical foundations for the propagation model, input signals, and observables presented in the main text. Unless otherwise stated, the notation follows that of the main manuscript.

\section{Discrete two-channel fiber model}
\label{sec:discrete-model}
This supplement derives the coupled propagation equations for two co-propagating optical fields in a single-mode optical fiber. Both channels are treated within the same fully quantum framework, without imposing assumptions regarding their relative field strengths or occupation numbers. The model assumes a scalar, polarization-preserving guided mode, an instantaneous electronic Kerr response, chromatic dispersion retained to second order, and Markovian linear loss. Raman and Brillouin contributions are neglected.

\subsection{Guided-mode expansion}

Let $\bm{x}=(\bm{x}_{\perp},z)$, where $z\in[0,L]$ is the longitudinal
coordinate and $\bm{x}_{\perp}$ denotes the transverse coordinates.  The two
co-propagating channels are centered at angular frequencies $\omega_1$ and
$\omega_2$. We discretize the fiber along the length into small cells. For longitudinal spectral component $m$, the longitudinal mode operators are denoted by $\hat a_m$ and
$\hat b_m$, respectively, and obey
\begin{equation}
    [\hat a_m,\hat a_{m'}^{\dagger}]=\delta_{mm'},
    \qquad
    [\hat b_m,\hat b_{m'}^{\dagger}]=\delta_{mm'},
    \qquad
    [\hat a_m,\hat b_{m'}^{\dagger}]=0.
    \label{eq:mode-commutators}
\end{equation}

For the discrete longitudinal Fourier representation used below, we introduce a uniform wave-number spacing $\Delta k=2\pi/L$. The full longitudinal wave number of the component indexed by $m$ in channel $j$ is
\begin{align}
k_m^{(j)} &= k_0^{(j)}+m\Delta k,
\qquad \Delta k=\frac{2\pi}{L},
\qquad j\in{1,2}.
\label{eq:S2}
\end{align}
Here, $k_0^{(j)}$ is the carrier wave number of channel $j$, and the integer $m$ labels the spectral component relative to that carrier.

Within the scalar-mode approximation, the positive-frequency electric displacement
operator for channel $j$ is written as \cite{raymer1991limits}
\begin{equation}
    \hat D_j^{(+)}(\bm{x})
    =
    \ii\sum_m
    \left(\frac{\epsilon_j\hbar v_j k_m^{(j)}}{2L}\right)^{1/2}
    \hat c_{j,m}
    u_j(\bm{x}_{\perp})
    \e^{\ii k_m^{(j)}z},
    \label{eq:displacement-expansion}
\end{equation}
where $\hat c_{1,m}=\hat a_m$, $\hat c_{2,m}=\hat b_m$,
$v_j=(\partial\omega/\partial k)_{k_0^{(j)}}$ is the group velocity, and
$\int\dd^2x_{\perp}|u_j(\bm{x}_{\perp})|^2=1$.  The full Hermitian field is
$\hat D_j=\hat D_j^{(+)}+\hat D_j^{(-)}$.  The effective Kerr coefficients
introduced below include the relevant transverse overlap integrals.  In
particular, the geometric-mean form used for cross-phase modulation assumes
the same polarization and transverse overlap for the two channels.

The fiber Hamiltonian is given by {\cite[Eqn. (3.2)]{drummond1987quantum}}

\begin{equation}
    \hat H_F
    =
    \sum_{j=1}^{2}\sum_m
    \hbar\omega_j(k_m^{(j)})
    \hat c_{j,m}^{\dagger}\hat c_{j,m}
    -\frac{\epsilon_0\chi^{(3)}}{4}
    \int\dd^3x\,:
    \left(
       \frac{\hat D_1\jn{(\bm{x})}}{\epsilon_1}
       +\frac{\hat D_2\jn{(\bm{x})}}{\epsilon_2}
    \right)^4: .
    \label{eq:fiber-hamiltonian}
\end{equation}
Normal ordering is indicated by colons. For a single channel, the field propagation is derived in \cite{b1, drummond2001quantum} including Kerr and Raman response.  Equation~\eqref{eq:fiber-hamiltonian}
uses the scalar effective susceptibility appropriate to the selected
polarization and guided mode \cite{drummond1987quantum}.

\subsection{Localized cell operators and Kerr interaction}

Divide the fiber into $M=2N+1$ cells of length $\Delta z=L/M$.  The localized
operators are the discrete Fourier transforms
\begin{align}
    \hat\alpha_l
    &=\frac{1}{\sqrt{M}}
      \sum_{m=-N}^{N}\hat a_m
      \e^{\ii m\Delta k l\Delta z},
    &
    \hat\beta_l
    &=\frac{1}{\sqrt{M}}
      \sum_{m=-N}^{N}\hat b_m
      \e^{\ii m\Delta k l\Delta z},
    \label{eq:localized-operators}
\end{align}
with $-N\leq l\leq N$.  They satisfy
$[\hat\alpha_l,\hat\alpha_{l'}^{\dagger}]
=[\hat\beta_l,\hat\beta_{l'}^{\dagger}]=\delta_{ll'}$.

After applying the rotating-wave approximation to $\hat{H}_F$ and retaining the self-phase-
and cross-phase-modulation terms, the nonlinear Hamiltonian becomes
\begin{align}
    \hat H_{\mathrm{NL}}
    ={}&
    -\hbar\tilde\chi_{1}
      \sum_l\hat\alpha_l^{\dagger 2}\hat\alpha_l^2
    -\hbar \tilde\chi_{2}
      \sum_l\hat\beta_l^{\dagger 2}\hat\beta_l^2
    \nonumber\\
    &-4\hbar\sqrt{\tilde\chi_{1} \tilde\chi_{2}}
      \sum_l
      \hat\alpha_l^{\dagger}\hat\alpha_l
      \hat\beta_l^{\dagger}\hat\beta_l .
    \label{eq:nonlinear-hamiltonian}
\end{align}
The factor of four in the last term follows from the multinomial expansion of the fourth-order field interaction. For identical transverse profiles and a cell quantization volume $\Delta V$, the self-Kerr strengths may be expressed as \cite[Eqn. (3.10)]{drummond1987quantum}
\begin{equation}
    \Tilde{\chi}_{j}
    =
    \frac{3\epsilon_0\chi^{(3)}\hbar
    v_{j}^{2}(k_0^{(j)})^{2}}
    {8\epsilon_{j}^{2}\Delta V}, \qquad j \in 1,2.
    \label{eq:discrete-kerr-strength}
\end{equation}
If the two channels have different transverse profiles, the three nonlinear
coefficients in Eq.~\eqref{eq:nonlinear-hamiltonian} must instead be evaluated
from their separate mode-overlap integrals.

In the linear part of the Hamiltonian in Eq.~\eqref{eq:fiber-hamiltonian}, the angular frequency $\omega$ and wave number $k$ are related through the channel-dependent dispersion relation $\omega^{(j)}(k)$. Expanding $\omega^{(j)}(k)$ about the carrier wave number $k_0^{(j)}$ gives rise to the group-velocity and group-velocity-dispersion (GVD) terms. Although the longitudinal wave numbers retained in the discrete representation are $k_m^{(j)}$, they sample this underlying dispersion relation. We therefore expand $\omega^{(j)}(k)$ about the carrier wave number $k_0^{(j)}$ and evaluate the expansion at the discrete wave numbers $k_m^{(j)}=k_0^{(j)}+q_m$:
\begin{align}
\omega^{(j)}\left(k_m^{(j)}\right)
&=
\omega^{(j)}\left(k_0^{(j)}+q_m\right)
=
\omega^{(j)}_0+v_jq_m+\frac{1}{2}\omega''^{(j)}q_m^2
+O\left(q_m^3\right),
\label{eq:dispersion-expansion}\\
q_m
&=
m\Delta k,
\qquad
\omega_0^{(j)}=\omega^{(j)}(k_0^{(j)}),
\nonumber\\
v_j
&=
\left.\frac{\mathrm{d}\omega^{(j)}}{\mathrm{d}k}\right|_{k_0^{(j)}},
\qquad
\omega''^{(j)}
=
\left.\frac{\mathrm{d}^2\omega''^{(j)}}{\mathrm{d}k^2}\right|_{k_0^{(j)}}.
\nonumber
\end{align}
Here, $q_m=m\Delta k$ is the discrete wave-number offset of spectral component $m$ from the carrier. In the present model, dispersion is retained only up to second order, while all terms of order $O(q_m^3)$ and higher are neglected.


In the localized basis, the linear coupling matrix for channel~$j$, in a
frame rotating at its carrier frequency is $\omega_{ll'}^{(j)}$. Following Ref.~\cite[Appendix~A]{b1}, we can write the linear coupling kernel as 
\begin{equation}
    \omega_{ll'}^{(j)}
    =
    \frac{1}{2N+1}\sum_{m=-N}^{N}
    \left[
        \omega^{(j)}\!\left(k_m^{(j)}\right)
        -\omega^{(j)}\!\left(k_0^{(j)}\right)
    \right]
    \e^{\ii m\Delta k(l-l')\Delta z},
    \label{eq:dispersion-matrix}
\end{equation}
where $k_0^{(j)}$ is the carrier wave number of channel~$j$. The subtraction
of the carrier frequency removes the uniform carrier-phase evolution.



\subsection{Linear-loss reservoirs}

To model distributed linear attenuation, we couple each localized guided-field operator to an independent continuum of environmental bosonic modes, following the reservoir construction of Ref.~[3]. The total Hamiltonian is
$\hat{H}_{\mathrm{tot}}=\hat{H}_{F}+\hat{H}_{R}+\hat{H}_{FR}$, where $\hat{H}_{F}$ is the fiber-field Hamiltonian introduced in Eq.~\eqref{eq:fiber-hamiltonian}, $\hat{H}_{R}$ describes the free reservoir modes, and $\hat{H}_{FR}$ describes the exchange of excitations between the guided fields and their reservoirs.

For each channel $j\in{1,2}$ and spatial cell $l$, let
$\hat{\gamma}_{j,l\mu}$ annihilate the $\mu$-th reservoir mode. These operators satisfy
\begin{equation}
\left[
\hat{\gamma}_{j,l\mu},
\hat{\gamma}_{k,l'\nu}^{\dagger}
\right]
=
\delta_{jk}\delta_{ll'}\delta_{\mu\nu},
\end{equation}
with all other commutators vanishing. In the frame rotating at the carrier frequency of channel $j$, the free-reservoir Hamiltonian is
\begin{equation}
\hat{H}_{R}
=
\hbar\sum_{j,l,\mu}
\omega_{\mu}^{(j)}
\hat{\gamma}_{j,l\mu}^{\dagger}
\hat{\gamma}_{j,l\mu},
\label{eq:reservoir-hamiltonian}
\end{equation}
where $(\omega_{\mu}^{(j)})$ denotes the detuning of reservoir mode $\mu$ from the carrier frequency of channel $j$.

The guided fields are coupled locally to the reservoirs through the interaction Hamiltonian
\begin{equation}
\hat{H}_{FR}
=
\hbar\sum_{j,l,\mu}
g_{j\mu}
\left(
\hat{c}_{j,l}^{\dagger}\hat{\gamma}_{j,l\mu}
+
\hat{c}_{j,l}\hat{\gamma}_{j,l\mu}^{\dagger}
\right),
\label{eq:field-reservoir-hamiltonian}
\end{equation}
where $\hat{c}_{1,l}=\hat{\alpha}_{l}$, $\hat{c}_{2,l}=\hat{\beta}_{l}$, and $g_{j\mu}$ is the field--reservoir coupling rate, taken to be real. The first term transfers an excitation from the reservoir to the guided field, whereas the second transfers an excitation from the guided field to the reservoir. 

The reservoirs are assumed independent for different channels and spatial cells. Consequently, tracing them out produces channel-local attenuation and the associated Langevin noise without introducing artificial coherent coupling or correlated loss between the two optical channels. Under the Markov approximation, this construction yields the damping and noise terms derived in Sec.~S2.3.

\section{\texorpdfstring{Positive-$P$}{Positive-P} equations}
\label{sec:positive-p}

\subsection{Phase-space representation}

The positive-$P$ representation expands the density operator as \cite{drummond1980generalised}
\begin{equation}
    \hat\rho(t)
    =\int P(\bm\lambda,t)
      \hat\Lambda(\bm\Psi,\bm\Psi^+)\,
      \dd\mu,
    \qquad
    \hat\Lambda
    =\frac{|\bm\Psi\rangle\langle(\bm\Psi^+)^*|}
    {\langle(\bm\Psi^+)^*|\bm\Psi\rangle},
    \label{eq:positive-p-expansion}
\end{equation}
where $\bm\lambda=(\bm\Psi,\bm\Psi^+)$ and $\dd\mu$ denotes integration over
the doubled complex phase space.  The variables $\alpha_l^+$ and $\beta_l^+$
are independent of $\alpha_l^*$ and $\beta_l^*$ on individual trajectories;
their ensemble moments reproduce normally ordered operator averages
\cite{drummond1980generalised,drummond1993quantum}.

Substituting the positive-$P$ expansion of the density operator into the interaction-picture Liouville equation, applying the identities in \cite[Eqs.~(10a)--(10d)]{b1} for the left and right action of the field operators on the kernel $\hat{\Lambda}$, and integrating by parts over the phase-space variables yields the standard Fokker--Planck equation

\begin{equation}
    \pd{P}{t}
    =-\sum_r\frac{\partial}{\partial\lambda_r}(A_rP)
     +\frac{1}{2}\sum_{r,s}
      \frac{\partial^2}{\partial\lambda_r\partial\lambda_s}
      (D_{rs}P).
    \label{eq:fokker-planck}
\end{equation}
Setting $\bm D=(D_{rs})_{r,s}$, where $r,s\in\{\alpha,\beta,\alpha^+,\beta^+\}$, we obtain for each cell the nonzero optical diffusion block
\begin{equation}
    \bm D=
    \begin{pmatrix}
      2\ii\Tilde{\chi}_1\alpha^2
      &4\ii\sqrt{\Tilde{\chi}_1\Tilde{\chi}_2}\alpha\beta&0&0\\
      4\ii\sqrt{\Tilde{\chi}_1\Tilde{\chi}_2}\alpha\beta
      &2\ii\Tilde{\chi}_2\beta^2&0&0\\
      0&0&-2\ii\Tilde{\chi}_1\alpha^{+2}
      &-4\ii\sqrt{\Tilde{\chi}_1\Tilde{\chi}_2}\alpha^+\beta^+\\
      0&0&-4\ii\sqrt{\Tilde{\chi}_1\Tilde{\chi}_2}\alpha^+\beta^+
      &-2\ii\Tilde{\chi}_2\beta^{+2}
    \end{pmatrix}.
    \label{eq:diffusion-matrix}
\end{equation}
Following Ref.~\cite[Eqns. (B1) and (B2)]{b1}, factorize $\bm D$ as $\bm D=\bm B\bm B^T$ to obtain the stochastic equations for the phase space variables which generate the individual trajectories needed for simulations. One convenient, nonunique factorization $\bm D=\bm B\bm B^T$ is
\begin{equation}
    \bm B=
    \begin{pmatrix}
      \sqrt{2\ii\Tilde{\chi}_1}\alpha&0&0&0\\
      \sqrt{8\ii\Tilde{\chi}_2}\beta&\sqrt{-6\ii\Tilde{\chi}_2}\beta&0&0\\
      0&0&\sqrt{-2\ii\Tilde{\chi}_1}\alpha^+&0\\
      0&0&\sqrt{-8\ii\Tilde{\chi}_2}\beta^+&
            \sqrt{6\ii\Tilde{\chi}_2}\beta^+
    \end{pmatrix}.
    \label{eq:diffusion-factorization}
\end{equation}
Direct multiplication verifies Eq.~\eqref{eq:diffusion-matrix}. This
factorization fixes the noise prefactors used throughout the remainder of the
supplement.

\subsection{Discrete It\^o equations}
Starting from the Liouville--von Neumann equation generated by the field--reservoir interaction Hamiltonian, the positive-$P$ operator correspondences convert the quantum evolution into the Fokker--Planck equation in Eq.~\eqref{eq:fokker-planck}. The coefficients of its first-order derivatives determine the deterministic drift, whereas those of its second-order derivatives form the diffusion matrix. Following the procedure described in \cite[Appendix B]{b1}, the factorization of the diffusion matrix given in Eq.~\eqref{eq:diffusion-factorization} maps the Fokker--Planck equation to an equivalent set of It\^{o} stochastic differential equations. Each column of the factorization matrix is associated with an independent real Gaussian noise source; when two phase-space variables couple to the same column, they are driven by the same noise realization, thereby reproducing the corresponding off-diagonal diffusion term. Applying this procedure gives the following discrete It\^{o} equations before elimination of the reservoir variables:

\begin{align}
    \pd{\alpha_l}{t}
    ={}&-\ii\sum_{l'}\omega_{ll'}^{(1)}\alpha_{l'}
      +2\ii\Tilde{\chi}_1\alpha_l^2\alpha_l^+
      +4\ii\sqrt{\Tilde{\chi}_1\Tilde{\chi}_2}
       \alpha_l\beta_l\beta_l^+
      -\ii\sum_\mu g_{1\mu}\gamma_{1,l\mu} \nonumber \\ 
      &+\sqrt{2\ii\Tilde{\chi}_1}\alpha_l\xi_l^{(1)},
    \label{eq:discrete-alpha}
    \\
    \pd{\beta_l}{t}
    ={}&-\ii\sum_{l'}\omega_{ll'}^{(2)}\beta_{l'}
      +2\ii\Tilde{\chi}_2\beta_l^2\beta_l^+
      +4\ii\sqrt{\Tilde{\chi}_1\Tilde{\chi}_2}
       \alpha_l\alpha_l^+\beta_l
      -\ii\sum_\mu 
      g_{2\mu}\gamma_{2,l\mu}
      \nonumber\\
    &+\sqrt{8\ii\Tilde{\chi}_2}\beta_l\xi_l^{(1)}
      +\sqrt{-6\ii\Tilde{\chi}_2}\beta_l\xi_l^{(2)},
    \label{eq:discrete-beta}
    \\
    \pd{\alpha_l^+}{t}
    ={}&\ii\sum_{l'}\omega_{ll'}^{(1)}\alpha_{l'}^+
      -2\ii\Tilde{\chi}_1\alpha_l^{+2}\alpha_l
      -4\ii\sqrt{\Tilde{\chi}_1\Tilde{\chi}_2}
       \alpha_l^+\beta_l^+\beta_l
      +\ii\sum_\mu g_{1\mu}\gamma_{1,l\mu}^+ \nonumber \\ 
      &+\sqrt{-2\ii\Tilde{\chi}_1}\alpha_l^+\xi_l^{+(1)},
    \label{eq:discrete-alpha-plus}
    \\
    \pd{\beta_l^+}{t}
    ={}&\ii\sum_{l'}\omega_{ll'}^{(2)}\beta_{l'}^+
      -2\ii\Tilde{\chi}_2\beta_l^{+2}\beta_l
      -4\ii\sqrt{\Tilde{\chi}_1\chi_2}
       \alpha_l^+\alpha_l\beta_l^+
      +\ii\sum_\mu g_{2\mu}\gamma_{2,l\mu}^+
      \nonumber\\
    &+\sqrt{-8\ii\Tilde{\chi}_2}\beta_l^+\xi_l^{+(1)}
      +\sqrt{6\ii\Tilde{\chi}_2}\beta_l^+\xi_l^{+(2)}
    \label{eq:discrete-beta-plus}
\end{align}
where we used Eq. \eqref{eq:dispersion-matrix}.
 The reservoir variables obey
\begin{equation}
    \pd{\gamma_{j,l\mu}}{t}
    =-\ii\omega_{j\mu}\gamma_{j,l\mu}
     -\ii g_{j\mu}c_{j,l},
    \qquad
    \pd{\gamma_{j,l\mu}^+}{t}
    =\ii\omega_{j\mu}\gamma_{j,l\mu}^+
     +\ii g_{j\mu}c_{j,l}^+.
    \label{eq:reservoir-sdes}
\end{equation}
The real noises have zero mean and correlations
\begin{align}
    \avg{\xi_l^{(r)}(t)\xi_{l'}^{(s)}(t')}
    &=\delta_{rs}\delta_{ll'}\delta(t-t'),
    \\
    \avg{\xi_l^{+(r)}(t)\xi_{l'}^{+(s)}(t')}
    &=\delta_{rs}\delta_{ll'}\delta(t-t'),
    \\
    \avg{\xi_l^{(r)}(t)\xi_{l'}^{+(s)}(t')}&=0.
    \label{eq:discrete-noise-correlations}
\end{align}
The same $\xi_l^{(1)}$ appears in Eqs.~\eqref{eq:discrete-alpha} and
\eqref{eq:discrete-beta}; this shared realization generates the required
off-diagonal diffusion element.

\subsection{Markov elimination of the loss reservoirs}

The formal solution of Eq.~\eqref{eq:reservoir-sdes} is
\begin{equation}
\gamma_{j,l\mu}(t)
=\gamma_{j,l\mu}(t_i)\e^{-\ii\omega_{j\mu}(t-t_i)}
-\ii g_{j\mu}\int_{t_i}^{t}\dd t',
\e^{-\ii\omega_{j\mu}(t-t')}c_{j,l}(t').
\label{eq:formal-reservoir-solution}
\end{equation}
The first term represents the freely evolving reservoir field, whereas the
integral describes the reservoir response to the guided field. For a broad,
weakly frequency-dependent reservoir, the response kernel becomes local in
time:
\begin{equation}
\sum_\mu g_{j\mu}^2
\e^{-\ii\omega_{j\mu}(t-t')}
\simeq\kappa_j\delta(t-t').
\label{eq:markov-kernel}
\end{equation}
Under this Markov approximation, the integral term produces local attenuation,
while the freely evolving reservoir field gives rise to a Langevin noise term.
Substitution into the field equations therefore gives
\begin{equation}
-\ii\sum_\mu g_{j\mu}\gamma_{j,l\mu}(t)
\longrightarrow
-\frac{\kappa_j}{2}c_{j,l}(t)+\Gamma_l^{(j)}(t).
\label{eq:reservoir-elimination}
\end{equation}
The factor $1/2$ follows from evaluating the delta function at the upper
integration limit. The statistics of $\Gamma_l^{(j)}$ are determined by the
initial reservoir state. For a thermal reservoir,
\begin{equation}
\avg{\Gamma_l^{(j)+}(t)\Gamma_{l'}^{(k)}(t')}
=\delta_{jk}\kappa_j n_{\mathrm{th},j}
\delta_{ll'}\delta(t-t').
\label{eq:reservoir-noise-correlation}
\end{equation}
At optical carrier frequencies and room temperature,
$n_{\mathrm{th},j}\simeq0$. Consequently, linear loss contributes the
deterministic attenuation term in normally ordered positive-$P$ equations,
while the thermal Langevin term may ordinarily be neglected. We retain it
symbolically below for completeness.

\section{Continuum propagation equations}
\label{sec:continuum}

\subsection{Continuum fields and nonlinear coefficients}

Take $\Delta z\rightarrow0$ while holding $z=l\Delta z$ fixed, and define
\begin{equation}
    \Phi_1(z,t)
    =\lim_{\substack{\Delta z\to0\\l\Delta z=z}}
      \alpha_l(t)\sqrt{\frac{v_1}{\Delta z}},
    \qquad
    \Phi_2(z,t)
    =\lim_{\substack{\Delta z\to0\\l\Delta z=z}}
      \beta_l(t)\sqrt{\frac{v_2}{\Delta z}}.
    \label{eq:continuum-fields}
\end{equation}
The $+$ fields are defined analogously.  With this normalization,
$\Phi_j^+\Phi_j$ is a photon flux and
$\hbar\omega_j\Phi_j^+\Phi_j$ is an optical power.

To make the continuum limit explicit and preserve the discrete diffusion
factorization, define the finite nonlinear coefficients 
\begin{equation}
    \chi_1=\lim_{\Delta z\to0}
        \frac{2\Tilde{\chi}_1\Delta z}{v_1^2},
    \qquad
    \chi_2=\lim_{\Delta z\to0}
        \frac{2\Tilde{\chi}_2\Delta z}{v_2^2}.
    \label{eq:continuum-kerr-coefficients}
\end{equation}
The discrete noises scale as
\begin{equation}
    \xi^{(r)}(z,t)
    =\lim_{\substack{\Delta z\to0\\l\Delta z=z}}
      \frac{\xi_l^{(r)}(t)}{\sqrt{\Delta z}},
    \qquad
    \avg{\xi^{(r)}(z,t)\xi^{(s)}(z',t')}
    =\delta_{rs}\delta(z-z')\delta(t-t'),
    \label{eq:continuum-noise}
\end{equation}
with identical relations for the independent $+$ noises.

Retaining dispersion through second order, Eqs.~\eqref{eq:discrete-alpha} and
\eqref{eq:discrete-beta} become
\begin{align}
    \pd{\Phi_1}{t}
    ={}&
    \left(-v_1\pd{}{z}
    +\frac{\ii\omega_1''}{2}\pdd{}{z}
    -\frac{\kappa_1}{2}\right)\Phi_1
    +\ii \chi_1v_1\Phi_1^2\Phi_1^+
    \nonumber\\
    &+2\ii\sqrt{\chi_1\chi_2}\,v_1\Phi_1\Phi_2\Phi_2^+
    +\sqrt{v_1}\,\Gamma^{(1)}
    +\sqrt{\ii \chi_1}\,v_1\Phi_1\xi^{(1)},
    \label{eq:continuum-phi1}
    \\
    \pd{\Phi_2}{t}
    ={}&
    \left(-v_2\pd{}{z}
    +\frac{\ii\omega_2''}{2}\pdd{}{z}
    -\frac{\kappa_2}{2}\right)\Phi_2
    + \ii \chi_2v_2\Phi_2^2\Phi_2^+
    \nonumber\\
    &+2\ii\sqrt{\chi_1\chi_2}\,v_2\Phi_1\Phi_1^+\Phi_2
    +\sqrt{v_2}\,\Gamma^{(2)}
    +\sqrt{4\ii \chi_2}\,v_2\Phi_2\xi^{(1)}
    +\sqrt{-3\ii \chi_2}\,v_2\Phi_2\xi^{(2)}.
    \label{eq:continuum-phi2}
\end{align}
Equations for $\Phi_1^+$ and $\Phi_2^+$ follow analogously from the doubled positive-$P$
sector.

\subsection{Retarded frame}

Choose the reference frame moving at the group velocity of channel~1,
\begin{equation}
    t'=t-\frac{z}{v_1},
    \qquad z'=z.
    \label{eq:retarded-coordinates}
\end{equation}
The derivatives transform according to
\begin{equation}
    \left.\pd{}{t}\right|_z=\left.\pd{}{t'}\right|_{z'},
    \qquad
    \left.\pd{}{z}\right|_t
    =\left.\pd{}{z'}\right|_{t'}
     -\frac{1}{v_1}\left.\pd{}{t'}\right|_{z'},
    \label{eq:first-derivative-transform}
\end{equation}
and hence
\begin{equation}
    \pdd{}{z}
    =\pdd{}{z'}-\frac{2}{v_1}
      \frac{\partial^2}{\partial z'\partial t'}
      +\frac{1}{v_1^2}\pdd{}{t'}.
    \label{eq:second-derivative-transform}
\end{equation}
Under the unidirectional slowly varying envelope approximation, the
$\partial_{z'}^2$ and $\partial_{z'}\partial_{t'}$ terms are of higher order
and are neglected.  Dividing the remaining equations by the corresponding
group velocities gives
\begin{align}
    \pd{\Phi_1}{z'}
    ={}&\left(
      \frac{\ii\omega_1''}{2v_1^3}\pdd{}{t'}
      -\frac{\kappa_1}{2v_1}\right)\Phi_1
      +\ii \chi_1\Phi_1^2\Phi_1^+ \nonumber\\
&+2\ii\sqrt{\chi_1\chi_2}\Phi_1\Phi_2\Phi_2^+
      +\frac{\Gamma^{(1)}}{\sqrt{v_1}}
      +\sqrt{\ii \chi_1}\Phi_1\xi^{(1)},
    \label{eq:retarded-phi1}
    \\
    \pd{\Phi_2}{z'}
    ={}&\left[\left(\frac{1}{v_1}-\frac{1}{v_2}\right)\pd{}{t'}
      +\frac{\ii\omega_2''}{2v_1^2v_2}\pdd{}{t'}
      -\frac{\kappa_2}{2v_2}\right]\Phi_2
      +\ii \chi_2\Phi_2^2\Phi_2^+
    \nonumber\\
    &+2\ii\sqrt{\chi_1\chi_2}\Phi_1\Phi_1^+\Phi_2
      +\frac{\Gamma^{(2)}}{\sqrt{v_2}}
      +\sqrt{4\ii \chi_2}\Phi_2\xi^{(1)}
      +\sqrt{-3\ii \chi_2}\Phi_2\xi^{(2)}.
    \label{eq:retarded-phi2}
\end{align}
The first-order temporal derivative in Eq.~\eqref{eq:retarded-phi2} is the
group-velocity mismatch, or walk-off, of channel~2 relative to channel~1.

\section{Dimensionless four-field model}
\label{sec:dimensionless-model}

\subsection{Scaling and coefficients}
The dispersion length is defined using the GVD coefficient $k''^{(j)}$, i.e., the curvature of the inverse dispersion relation $k^{(j)}(\omega)$. Since $k^{(j)}(\omega)$ and $\omega^{(j)}(k)$ are inverse functions and $\partial k/\partial\omega=1/v_j$, differentiation gives $k''^{(j)}=-\omega''^{(j)}/v_j^3$. Introducing a characteristic time $t_0$, we therefore define
\begin{align}
k''^{(j)}
&\equiv
\left.
\frac{\partial^2k^{(j)}}{\partial\omega^2}
\right|_{\omega_0''^{(j)}}
=
-\frac{\omega''^{(j)}}{v_j^3},
\qquad
z_0
=
\frac{t_0^2}{|k''^{(1)}|}.
\label{eq:dispersion-length}
\end{align}
Here, $\omega''^{(j)}$ is the curvature of $\omega^{(j)}(k)$ at $k_0^{(j)}$, as defined in Eq.~\eqref{eq:dispersion-expansion}. \\

Define the dimensionless coordinates and fields by
\begin{equation}
    \zeta=\frac{z'}{z_0},
    \qquad
    \tau=\frac{t'}{t_0},
    \qquad
    \Phi_j=\sqrt{\frac{P_j}{\hbar\omega_j}}\,u_j,
    \qquad
    \Phi_j^+=\sqrt{\frac{P_j}{\hbar\omega_j}}\,u_j^+.
    \label{eq:dimensionless-variables}
\end{equation}
Here $P_j$ is the peak power of the signal.
The linear dimensionless parameters are
\begin{align}
    W&=\left(\frac{1}{v_1}-\frac{1}{v_2}\right)\frac{z_0}{t_0},
    \qquad
    s_1=\operatorname{sgn}(k_1''),
    \qquad
    D_2=\frac{k_2''}{|k_1''|}
          \left(\frac{v_2}{v_1}\right)^2,
    \label{eq:linear-dimensionless-parameters}
\end{align}
where the dispersion operators are
$-\ii s_1\partial_\tau^2/2$ and
$-\ii D_2\partial_\tau^2/2$.  Thus the channel-1 term is
$+\ii\partial_\tau^2/2$ only for anomalous dispersion, $k_1''<0$.

The nonlinear coefficients are
\begin{align}
    G_{11}&=\frac{\chi_1z_0P_1}{\hbar\omega_1},
    &
    G_{12}&=\frac{2\sqrt{\chi_1\chi_2}\,z_0P_2}{\hbar\omega_2},
    \nonumber\\
    G_{22}&=\frac{\chi_2z_0P_2}{\hbar\omega_2},
    &
    G_{21}&=\frac{2\sqrt{\chi_1\chi_2}\,z_0P_1}{\hbar\omega_1}.
    \label{eq:nonlinear-dimensionless-parameters}
\end{align}
The dimensionless white noises are
\begin{equation}
    \eta_r(\zeta,\tau)=\sqrt{z_0t_0}\,\xi^{(r)}(z',t'),
    \qquad
    \avg{\eta_r(\zeta,\tau)\eta_s(\zeta',\tau')}
    =\delta_{rs}\delta(\zeta-\zeta')\delta(\tau-\tau'),
    \label{eq:dimensionless-noise}
\end{equation}
with independent $\eta_r^+$ satisfying the same autocorrelations and
$\avg{\eta_r\eta_s^+}=0$.

For completeness, the scaled thermal-reservoir terms are
\begin{equation}
    R_j=\Gamma^{(j)}
    \sqrt{\frac{\hbar\omega_jz_0^2}{P_jv_j}},
    \qquad j\in\{1,2\},
    \label{eq:scaled-reservoir-noise}
\end{equation}
and similarly for $R_j^+$.  They are highly suppressed as
$n_{\mathrm{th},j}\simeq0$.

\subsection{Common fiber loss}

A power attenuation of $0.2~\mathrm{dB/km}$ corresponds to the field-amplitude
attenuation coefficient
\begin{equation}
    \alpha_{\mathrm f}
    =\frac{\ln(10)}{20}
      \left(0.2\times10^{-3}\right)
    =\ln(10)\times10^{-5}\ \mathrm{m}^{-1}.
    \label{eq:field-loss}
\end{equation}
Assuming the same loss in both channels gives
\begin{equation}
    \frac{\kappa_1}{2v_1}
    =\frac{\kappa_2}{2v_2}
    =\alpha_{\mathrm f},
    \qquad
    \bar{\alpha}_{\mathrm f}=\alpha_{\mathrm f}z_0,
    \label{eq:common-loss}
\end{equation}
where $z_0$ is expressed in meters.  Notice that the $1/20$ conversion is
appropriate for a field amplitude; the corresponding optical power decays as
$\exp(-2\alpha_{\mathrm f}z)$.

\subsection{Final coupled equations}
With the definitions above, the four positive-$P$ propagation equations are
\begin{align}
    \pd{u_1}{\zeta}
    ={}&\left(
      -\frac{\ii s_1}{2}\pdd{}{\tau}-\bar{\alpha}_{\mathrm f}\right)u_1
      +\ii G_{11}u_1^2u_1^+
      +\ii G_{12}u_1u_2u_2^+
      +R_1
    \nonumber\\
    &+\sqrt{\frac{\ii \chi_1z_0}{t_0}}\,u_1\eta_1,
    \label{eq:final-u1}
    \\
    \pd{u_1^+}{\zeta}
    ={}&\left(
    \frac{\ii s_1}{2}\pdd{}{\tau}-\bar{\alpha}_{\mathrm f}\right)u_1^+
      -\ii G_{11}(u_1^+)^2u_1
      -\ii G_{12}u_1^+u_2u_2^+
      +R_1^+
    \nonumber\\
    &+\sqrt{\frac{-\ii \chi_1z_0}{t_0}}\,u_1^+\eta_1^+,
    \label{eq:final-u1-plus}
    \\
    \pd{u_2}{\zeta}
    ={}&\left(W\pd{}{\tau}
      -\frac{\ii D_2}{2}\pdd{}{\tau}-\bar{\alpha}_{\mathrm f}\right)u_2
      +\ii G_{22}u_2^2u_2^+
      +\ii G_{21}u_1u_1^+u_2
      +R_2
    \nonumber\\
    &+\sqrt{\frac{4\ii \chi_2z_0}{t_0}}\,u_2\eta_1
      +\sqrt{\frac{-3\ii \chi_2z_0}{t_0}}\,u_2\eta_2,
    \label{eq:final-u2}
    \\
    \pd{u_2^+}{\zeta}
    ={}&\left(W\pd{}{\tau}
      +\frac{\ii D_2}{2}\pdd{}{\tau}-\bar{\alpha}_{\mathrm f}\right)u_2^+
      -\ii G_{22}(u_2^+)^2u_2
      -\ii G_{21}u_1u_1^+u_2^+
      +R_2^+
    \nonumber\\
    &+\sqrt{\frac{-4\ii \chi_2z_0}{t_0}}\,u_2^+\eta_1^+
      +\sqrt{\frac{3\ii \chi_2z_0}{t_0}}\,u_2^+\eta_2^+.
    \label{eq:final-u2-plus}
\end{align}
The four equations must be propagated together because $u_j^+$ is not the
trajectory-wise complex conjugate of $u_j$.  Physical normally ordered moments
are recovered as stochastic ensemble averages; for example,
\begin{equation}
    \avg{:\hat n_1^m\hat n_2^n:}
    \longleftrightarrow
    \avg{(u_1^+u_1)^m(u_2^+u_2)^n}_{\mathrm{stoch}},
    \label{eq:observable-mapping}
\end{equation}
up to the power-normalization factors in
Eq.~\eqref{eq:dimensionless-variables}.  Equations
\eqref{eq:final-u1}--\eqref{eq:final-u2-plus} are the dimensionless model used
for the co-propagation analysis.

\subsection{Stochastic convention and numerical implementation}

The positive-$P$ Fokker--Planck equation yields stochastic equations in the It\^{o} convention. Because the semi-implicit midpoint scheme used for numerical propagation is formulated in the Stratonovich convention, the standard It\^{o}-to-Stratonovich conversion is applied. Let $\lambda_r$ denote the doubled phase-space variables, $B_{rs}$ the diffusion-factorization matrix satisfying $D=BB^{\mathsf T}$, and $A_r$ the It\^{o} drift. The corresponding Stratonovich drift is
\begin{equation}
A_r^{(\mathrm S)}
=
A_r
-\frac{1}{2}
\sum_{k,s}
B_{ks}
\frac{\partial B_{rs}}{\partial\lambda_k}.
\label{eq:ito-to-stratonovich}
\end{equation}

Applying Eq.~\eqref{eq:ito-to-stratonovich} to the diffusion factorization used in Eqs.~\eqref{eq:discrete-alpha}--\eqref{eq:discrete-beta-plus} gives
\begin{equation}
\boldsymbol{A}^{(\mathrm S)}
=
\boldsymbol{A}
+
\left(
-\ii\widetilde{\chi}_1\alpha_l,,
-\ii\widetilde{\chi}_2\beta_l,,
+\ii\widetilde{\chi}_1\alpha_l^+,,
+\ii\widetilde{\chi}_2\beta_l^+
\right)^{\mathsf T}.
\label{eq:stratonovich-correction}
\end{equation}
The additive reservoir-noise terms do not contribute to this conversion because their amplitudes are independent of the phase-space variables.

The deterministic parts of Eqs.~\eqref{eq:final-u1}--\eqref{eq:final-u2-plus} are obtained from the It\^{o} drifts. Carrying the corrections in Eq.~\eqref{eq:stratonovich-correction} through the continuum-field transformation, retarded-frame transformation, and dimensionless scaling gives the additional drift terms required by the Stratonovich midpoint scheme:
\begin{align}
\Delta A_1
&=
-\frac{\ii}{2}
\frac{\chi_1 z_0}{\Delta t}\,u_1,
&
\Delta A_1^+
&=
+\frac{\ii}{2}
\frac{\chi_1 z_0}{\Delta t}\,u_1^+,
\\
\Delta A_2
&=
-\frac{\ii}{2}
\frac{\chi_2 z_0}{\Delta t}\,u_2,
&
\Delta A_2^+
&=
+\frac{\ii}{2}
\frac{\chi_2 z_0}{\Delta t}\,u_2^+.
\label{eq:dimensionless-stratonovich-corrections}
\end{align}

Here, $\Delta t=t_0\Delta\tau$ is the physical spacing of the
temporal grid used to discretize the delta-correlated noise in the
retarded-time coordinate. In particular, the temporal Dirac delta is
represented on the grid as
$\delta(t-t')\rightarrow\delta_{nn'}/\Delta t$.
With
$A^{(\mathrm{S})}=A+\Delta A$,
these drift corrections make the semi-implicit midpoint
(Stratonovich) implementation equivalent to the underlying
It\^{o} positive-$P$ equations. They are discretization corrections
and do not represent an additional physical propagation mechanism.

\section{Scope of the model}
\label{sec:scope}

The model is formulated under assumptions appropriate to the propagation regime considered: (i) each carrier occupies a single scalar guided mode; (ii) the Kerr response is instantaneous and electronic; (iii) chromatic dispersion is retained through $k''^{(j)}$; (iv) propagation is unidirectional and described by slowly varying envelopes; and (v) linear attenuation is represented by independent Markov reservoirs. Within this framework, the model consistently describes loss, group-velocity mismatch, second-order dispersion, self- and cross-phase modulation, and the associated quantum-stochastic dynamics. Polarization coupling, higher-order dispersion, self-steepening, delayed Raman response, and Brillouin interactions may be incorporated through corresponding extensions of the Hamiltonian and stochastic-noise model.


\section{Quadrature fluctuation spectrum}
\label{sec:qfs}
Following the balanced-homodyne photocurrent-correlation approach
used for optical solitons in Ref.~\cite{drummond1989time} and the
pulsed-field formulation presented in Sec.~IV of
Ref.~\cite{werner1997pulsed}, we define the frequency-resolved
quadrature-fluctuation spectrum from the Fourier-domain correlation
of the mean-subtracted balanced-homodyne difference photocurrent.
Normal ordering separates the shot-noise contribution from the
field-dependent fluctuation contribution, allowing the latter to be
evaluated directly using positive-$P$ stochastic fields. We apply
this formulation to the output of the propagation model in
Eqs.~\eqref{eq:final-u1}-\eqref{eq:final-u2-plus}. We use the notation and normalization introduced in the propagation supplement: $\Phi_j$ and $\Phi_j^+$ denote the continuum positive-$P$ fields, while $u_j$ and $u_j^+$ are their dimensionless counterparts. Because the plus field is statistically independent of the complex-conjugate field on each stochastic trajectory, the positive- and negative-frequency components entering the normally ordered spectrum must be evaluated directly from the corresponding stochastic fields. This distinction is essential for correctly reconstructing normally ordered quadrature correlations from the stochastic ensemble.

\subsection{Propagated field at the detector}
\label{subsec:qfs-output-field}

Let channel $j\in\{1,2\}$ be measured at a fixed output distance
$z'=z_{\out}$, or equivalently $\zeta=\zeta_{\out}$.  We suppress that fixed
propagation coordinate and write the signal operator as
$\hat\Phi_j(t')$.  It obeys
\begin{equation}
    \left[
        \hat\Phi_j(t'),\hat\Phi_k^\dagger(t'')
    \right]
    =\delta_{jk}\delta(t'-t''),
    \label{eq:qfs-field-commutator}
\end{equation}
so that $\hat\Phi_j^\dagger\hat\Phi_j$ is a photon flux.  The scaling used in
Eq.~\eqref{eq:dimensionless-variables} is
\begin{equation}
    t'=t_0\tau,
    \qquad
    \Phi_j(t')
      =\sqrt{\frac{P_j}{\hbar\omega_j}}\,u_j(\tau),
    \qquad
    \Phi_j^+(t')
      =\sqrt{\frac{P_j}{\hbar\omega_j}}\,u_j^+(\tau).
    \label{eq:qfs-field-scaling}
\end{equation}
The positive-$P$ correspondence for a normally ordered functional is
\begin{equation}
    \avg{\normal{F[\hat\Phi_j^\dagger,\hat\Phi_j]}}
    =
    \stavg{F[\Phi_j^+,\Phi_j]}.
    \label{eq:qfs-positive-p-correspondence}
\end{equation}
Equation~\eqref{eq:qfs-positive-p-correspondence} is the same ordering rule
used for the moments in Eq.~\eqref{eq:observable-mapping}.

\subsection{Balanced-homodyne observable}
\label{subsec:qfs-homodyne}

Let $\hat\Phi_{\LO}(t')$ denote an independent local-oscillator (LO) mode.  At
a lossless balanced beam splitter, choose the output fields
\begin{equation}
    \hat\Phi_c(t')
      =\frac{\hat\Phi_j(t')+
              \mathrm{e}^{\ii\theta}\hat\Phi_{\LO}(t')}{\sqrt{2}},
    \qquad
    \hat\Phi_d(t')
      =\frac{\hat\Phi_j(t')-
              \mathrm{e}^{\ii\theta}\hat\Phi_{\LO}(t')}{\sqrt{2}}.
    \label{eq:qfs-beamsplitter}
\end{equation}
The photon-flux difference is
\begin{align}
    \hat X_{j,\theta}(t')
    &\equiv
      \hat\Phi_c^\dagger(t')\hat\Phi_c(t')
      -\hat\Phi_d^\dagger(t')\hat\Phi_d(t')
      \nonumber\\
    &=\mathrm{e}^{\ii\theta}
      \hat\Phi_j^\dagger(t')\hat\Phi_{\LO}(t')
      +\mathrm{e}^{-\ii\theta}
      \hat\Phi_{\LO}^\dagger(t')\hat\Phi_j(t').
    \label{eq:qfs-homodyne-observable}
\end{align}
The LO phase $\theta$ selects the quadrature, and its temporal envelope selects
the temporal mode.  We define the mean-subtracted photocurrent fluctuation as
\begin{equation}
    \delta\hat X_{j,\theta}(t')
    =\hat X_{j,\theta}(t')
     -\avg{\hat X_{j,\theta}(t')}.
    \label{eq:qfs-connected-fluctuation}
\end{equation}
The corresponding two-time covariance, $\langle \delta \hat{X}_{j,\theta}(t') \delta \hat{X}_{j,\theta}(t'') \rangle$, removes the coherent mean-field contribution and characterizes the photocurrent fluctuations.

\subsection{Finite-window spectrum and shot-noise normalization}
\label{subsec:qfs-spectrum-definition}

For a detection window $\mathcal W=[-T/2,T/2]$, define
\begin{equation}
    \delta\hat X_{j,\theta}(\omega)
    =\frac{1}{\sqrt{2\pi}}
      \int_{\mathcal W}\dd t'\,
      \mathrm{e}^{-\ii\omega t'}
      \delta\hat X_{j,\theta}(t').
    \label{eq:qfs-fourier-transform}
\end{equation}
Here $\omega$ is the electronic analysis angular frequency relative to the
optical carrier.  The minus sign in
Eq.~\eqref{eq:qfs-fourier-transform} matches the FFT convention used for the
propagation calculation.  Hermiticity gives
$\delta\hat X_{j,\theta}(-\omega)
=\delta\hat X_{j,\theta}^\dagger(\omega)$.

The mean photon number incident on both detectors during the same window is
\begin{equation}
    N_{\det}
    =\int_{\mathcal W}\dd t'\,
      \avg{
        \hat\Phi_j^\dagger(t')\hat\Phi_j(t')
        +\hat\Phi_{\LO}^\dagger(t')\hat\Phi_{\LO}(t')
      }.
    \label{eq:qfs-detected-photon-number}
\end{equation}
Using Eq.~\eqref{eq:qfs-field-commutator} to normal order the two-time
photocurrent product \cite{werner1997pulsed, drummond1989time} gives
\begin{align}
    \avg{
      \delta\hat X_{j,\theta}(-\omega)
      \delta\hat X_{j,\theta}(\omega)
    }
    & =\frac{N_{\det}}{2\pi}
    \nonumber\\
    &\quad+
    \avg{
      \normal{
        \delta\hat X_{j,\theta}(-\omega)
        \delta\hat X_{j,\theta}(\omega)
      }
    }.
    \label{eq:qfs-shot-noise-decomposition}
\end{align}
Therefore, the quadrature noise spectrum is given by
\begin{equation}
    S^{(N)}_{j,\theta}(\omega)
    =\frac{2\pi}{N_{\det}}
      \avg{
        \normal{
          \delta\hat X_{j,\theta}(-\omega)
          \delta\hat X_{j,\theta}(\omega)
        }
      },
    \label{eq:qfs-normal-spectrum}
\end{equation}

The above formula is used to calculate the quadrature fluctuation of an optical soliton propagating in a fiber \cite{drummond1987quantum}.
A coherent state has $V_{j,\theta}=1$ and $S^{(N)}_{j,\theta}=0$.
Quadrature squeezing occurs when
\begin{equation}
    V_{j,\theta}(\omega)<1,
    \qquad\text{equivalently}\qquad
    S^{(N)}_{j,\theta}(\omega)<0.
    \label{eq:qfs-squeezing-condition}
\end{equation}
The squeezing level in decibels is
$10\log_{10}V_{j,\theta}(\omega)$.

\subsection{Positive-\texorpdfstring{$P$}{P} estimator}
\label{subsec:qfs-positive-p-estimator}

On trajectory $r$, the normally ordered stochastic variable corresponding to
Eq.~\eqref{eq:qfs-homodyne-observable} is
\begin{equation}
    X_{j,\theta}^{(r)}(t')
    =\mathrm{e}^{\ii\theta}
      \Phi_j^{+(r)}(t')\Phi_{\LO}^{(r)}(t')
     +\mathrm{e}^{-\ii\theta}
      \Phi_{\LO}^{+(r)}(t')\Phi_j^{(r)}(t').
    \label{eq:qfs-positive-p-current}
\end{equation}
For each stochastic trajectory, define the mean-subtracted photocurrent and its Fourier transform by
\begin{equation}
    \delta X_{j,\theta}^{(r)}(t')
      =X_{j,\theta}^{(r)}(t')
       -\stavg{X_{j,\theta}(t')},
    \qquad
    \delta X_{j,\theta}^{(r)}(\omega)
      =\frac{1}{\sqrt{2\pi}}
       \int_{\mathcal W}\dd t'\,
       \mathrm{e}^{-\ii\omega t'}
       \delta X_{j,\theta}^{(r)}(t').
    \label{eq:qfs-positive-p-connected-fourier}
\end{equation}
The detected photon number is evaluated as
\begin{equation}
    N_{\det}^{(P)}
    =\int_{\mathcal W}\dd t'\,
      \stavg{
        \Phi_j^+(t')\Phi_j(t')
        +\Phi_{\LO}^+(t')\Phi_{\LO}(t')
      }.
    \label{eq:qfs-positive-p-detected-number}
\end{equation}
Equations~\eqref{eq:qfs-positive-p-correspondence} and
\eqref{eq:qfs-normal-spectrum} then give the directly sampled expressions
\begin{align}
    S^{(N,P)}_{j,\theta}(\omega)
    &=\frac{2\pi}{N_{\det}^{(P)}}
      \stavg{
        \delta X_{j,\theta}(-\omega)
        \delta X_{j,\theta}(\omega)
      },
    \label{eq:qfs-positive-p-normal-spectrum}\\
    V^{(P)}_{j,\theta}(\omega)
    &=1+S^{(N,P)}_{j,\theta}(\omega).
    \label{eq:qfs-positive-p-total-spectrum}
\end{align}
The explicit $+1$ in Eq.~\eqref{eq:qfs-positive-p-total-spectrum} restores
the commutator, or shot-noise, contribution that is absent from normally
ordered positive-$P$ averages.

The variables $\Phi_j^+$ and $\Phi_j$ are independent on each trajectory.
Consequently,
$\delta X_{j,\theta}^{(r)}(-\omega)$ must be constructed from the negative
frequency bin and must not be replaced by
$[\delta X_{j,\theta}^{(r)}(\omega)]^*$ or by an absolute square.  The
ensemble-averaged spectrum is real in the infinite-sample limit; any residual
imaginary part is a sampling diagnostic.

\subsection{Strong classical pulsed LO}
\label{subsec:qfs-strong-lo}

For a classical pulsed LO, write
\begin{equation}
    \Phi_{\LO}(t')=\alpha_{\LO}f_{\LO}(t'),
    \qquad
    \Phi_{\LO}^+(t')=\alpha_{\LO}^*f_{\LO}^*(t'),
    \qquad
    \int_{\mathcal W}\dd t'\,|f_{\LO}(t')|^2=1.
    \label{eq:qfs-classical-lo}
\end{equation}
The constant phase of $\alpha_{\LO}$ can be absorbed into $\theta$, so below
we take $\alpha_{\LO}=|\alpha_{\LO}|$.  Then
$N_{\LO}=|\alpha_{\LO}|^2$, and in the strong-LO limit
$N_{\LO}\gg N_j$ one has $N_{\det}\simeq N_{\LO}$.  Define the LO-normalized
signal quadrature
\begin{equation}
    x_{j,\theta}^{(r)}(t')
    =\mathrm{e}^{\ii\theta}
      \Phi_j^{+(r)}(t')f_{\LO}(t')
     +\mathrm{e}^{-\ii\theta}
      f_{\LO}^*(t')\Phi_j^{(r)}(t'),
    \label{eq:qfs-lo-normalized-quadrature}
\end{equation}
so that $X_{j,\theta}^{(r)}=|\alpha_{\LO}|x_{j,\theta}^{(r)}$.  The LO
amplitude cancels between numerator and denominator, giving
\begin{equation}
    \boxed{
    V^{(P)}_{j,\theta}(\omega)
    \simeq
    1+2\pi\,
      \stavg{
        \delta x_{j,\theta}(-\omega)
        \delta x_{j,\theta}(\omega)
      }.}
    \label{eq:qfs-strong-lo-spectrum}
\end{equation}

```latex
\section{Numerical convergence of the quadrature spectra}
\label{sec:quadrature-convergence}

To verify that the small quadrature-fluctuation spectra are not artifacts of finite stochastic sampling, we repeated the calculation at an average classical launch power of
$P_{2}^{(\mathrm{avg})}=40\,\mathrm{mW}$ and at the fiber output $z=100\,\mathrm{km}$ using successively enlarged ensembles,
$N_{\mathrm{st}}=2^{8}$, $2^{10}$, $2^{12}$, and $2^{14}$.
All physical parameters and numerical discretization settings were kept unchanged.

\begin{figure}[!t]
    \centering
    \includegraphics[width=0.98\linewidth]{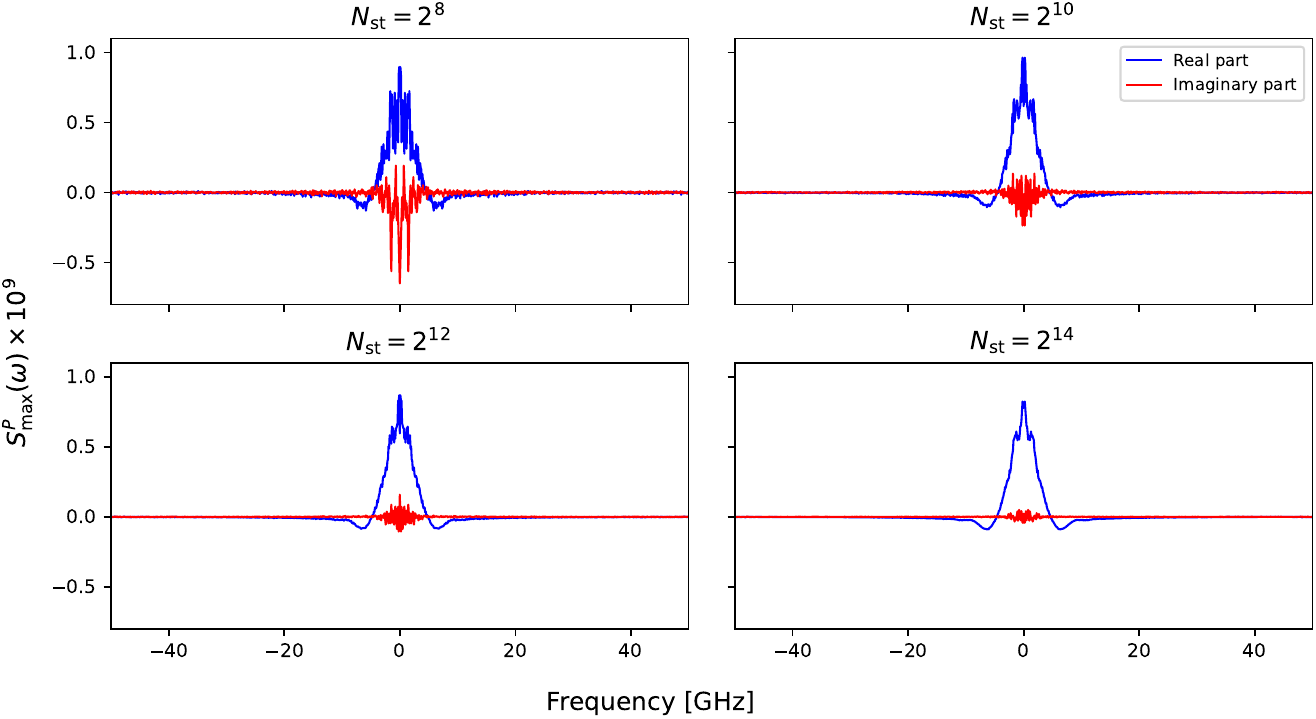}
    \caption{Ensemble-size convergence of the maximum normally ordered quadrature-fluctuation spectrum $S_{\max}^{P}(\omega)$ at $P_{2}^{(\mathrm{avg})}=40\,\mathrm{mW}$ and $z=100\,\mathrm{km}$. The blue and red curves show the real and residual imaginary parts, respectively, for increasing numbers of positive-$P$ trajectories. The horizontal axis is the analysis frequency $f=\omega/(2\pi)$. All panels use the same vertical scale.}
    \label{fig:quadrature-convergence-max}
\end{figure}

\begin{figure}[!t]
    \centering
    \includegraphics[width=0.98\linewidth]{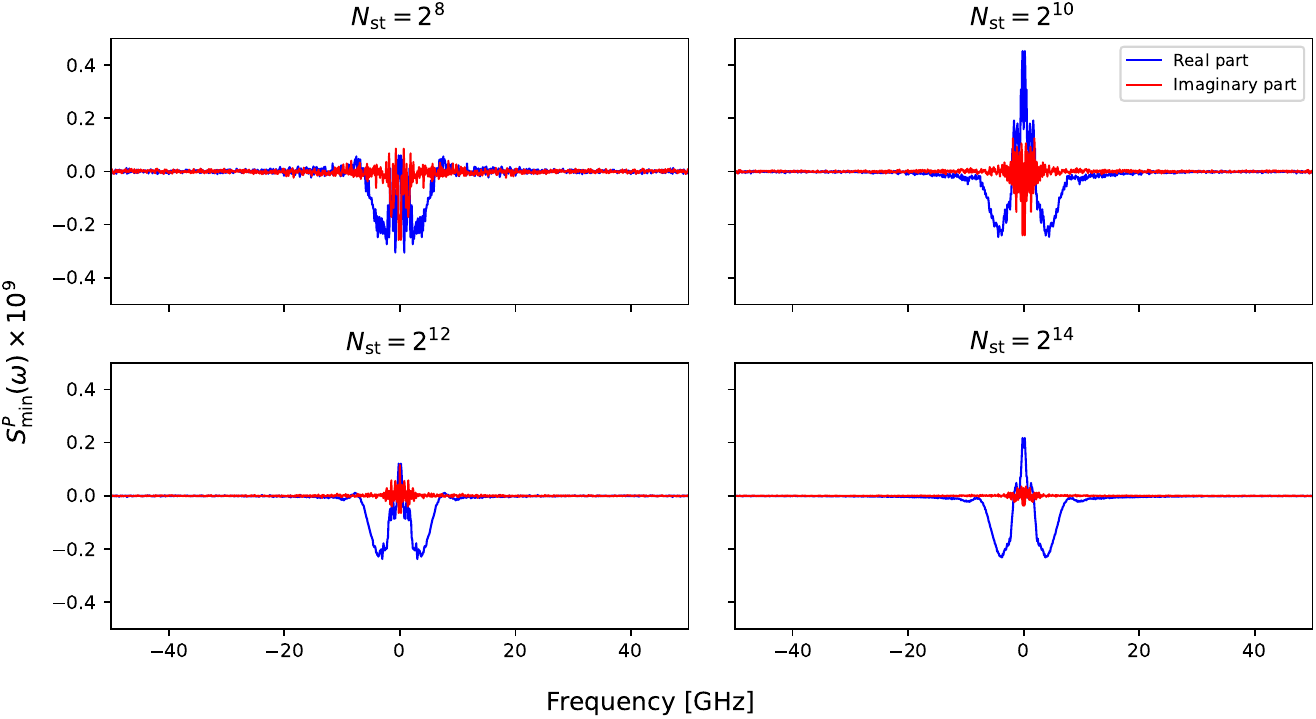}
    \caption{Ensemble-size convergence of the minimum normally ordered quadrature-fluctuation spectrum $S_{\min}^{P}(\omega)$ under the same conditions as Fig.~\ref{fig:quadrature-convergence-max}. The persistence of the real spectral structure together with the suppression of the residual imaginary component demonstrates convergence with increasing stochastic ensemble size.}
    \label{fig:quadrature-convergence-min}
\end{figure}

As discussed in Sec.~\ref{subsec:qfs-positive-p-estimator}, the positive-$P$ estimator of the quadrature spectrum is real in the infinite-ensemble limit. Consequently, a residual imaginary component provides a direct diagnostic of finite-sampling error. Figures~\ref{fig:quadrature-convergence-max} and \ref{fig:quadrature-convergence-min} show the real and imaginary parts of the maximum- and minimum-quadrature spectra, respectively. At small ensemble sizes, both spectra exhibit appreciable stochastic irregularity and residual imaginary components. As $N_{\mathrm{st}}$ increases, the real spectra approach stable frequency-dependent profiles, while the imaginary components are progressively suppressed. In particular, the spectral bandwidth and the positions of the principal extrema are preserved between $N_{\mathrm{st}}=2^{12}$ and $2^{14}$, with substantially smaller sampling fluctuations at the latter ensemble size.

The real spectral features remain of order $10^{-9}$ as the ensemble size is increased, whereas the finite-sampling imaginary contribution decreases. The reported spectral structure is therefore statistically resolved rather than dominated by stochastic sampling noise. Accordingly, all quadrature-fluctuation spectra reported in the main text were evaluated using $N_{\mathrm{st}}=2^{14}$ trajectories.

\section{Quadrature Amplitude Modulation}
\label{sec:qam}

\begin{figure*}[ht]
     \centering
     \includegraphics[width=0.98\linewidth]{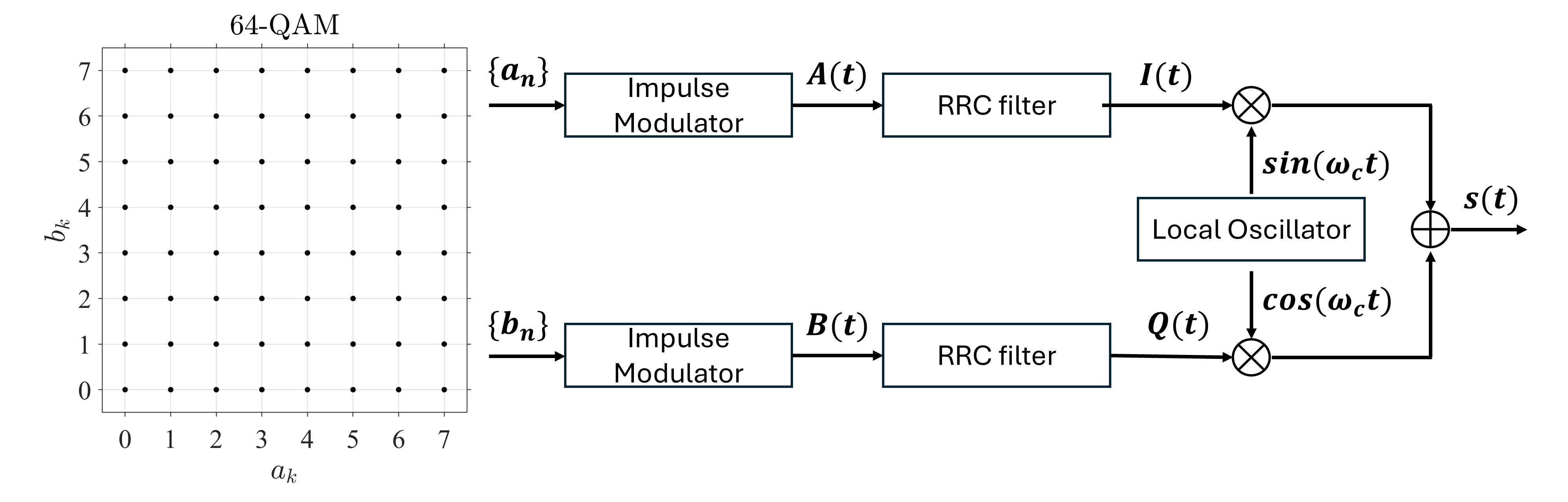}
    \caption{Block diagram of \gls{qam} transmitter}
     \label{dia_qam}
 \end{figure*}

\gls{qam} is a widely utilized digital modulation technique in modern telecommunication systems due to its high spectral efficiency and robustness to noise. In an $M$-\gls{qam} scheme, each transmitted symbol encodes $\log_2(M)$ bits and is represented as a point in a two-dimensional signal constellation. The $k^{\text{th}}$ symbol can be expressed as a complex number
\[
c_k = a_k + i b_k,
\]
where $a_k$ and $b_k$ denote the in-phase (I) and quadrature (Q) components, respectively. 
A random symbol sequence ${c_k}$, with each $c_k$ drawn from the $M$-QAM constellation, is generated at the symbol rate $f_{\mathrm{sym}}$. Each component $a_k$ and $b_k$ carries $\log_2(M)/2$ bits of information and is subsequently processed by an impulse modulator, as illustrated in Fig.~\ref{dia_qam}.

Let $f_{\mathrm{sym}}$ be the symbol rate in baud, then $T_{\mathrm{sym}} = 1/f_{\mathrm{sym}}$ is the symbol period of a QAM symbol. The corresponding analog impulses for the I and Q components are
\begin{subequations}
\begin{align}
A(t) &= \sum_k a_k \, \delta(t - k\ T_s), \\
B(t) &= \sum_k b_k \, \delta(t - k\ T_s),
\end{align}
\end{subequations}
To optimize spectral efficiency and mitigate intersymbol interference, these impulses are shaped using a \gls{rrc} filter:
\begin{equation}
h_\text{RRC}(t) =
\begin{cases}
f_{\mathrm{sym}} \left[ 1 + \beta \left( \dfrac{4}{\pi} - 1 \right) \right], & t = 0, \\
f_{\mathrm{sym}} \dfrac{\beta}{\sqrt{2}} \left[ \left( 1 + \dfrac{2}{\pi} \right) \sin\left(\dfrac{\pi}{4\beta}\right) + \left( 1 - \dfrac{2}{\pi} \right) \cos\left(\dfrac{\pi}{4\beta}\right) \right], & t = \pm \dfrac{1}{4 \beta f_{\mathrm{sym}}}, \\
\dfrac{\sin\left[\pi t f_{\mathrm{sym}} (1 - \beta)\right] + 4 \beta t f_{\mathrm{sym}} \cos\left[\pi f_{\mathrm{sym}} (1 + \beta)\right]}{\pi t \left[ 1 - (4 \beta t f_{\mathrm{sym}})^2 \right]}, & \text{otherwise},
\end{cases}
\end{equation}

where $\beta$ is the roll-off factor controlling the trade-off between spectral efficiency and intersymbol interference suppression.

The filtered I and Q components are obtained as
\begin{subequations}
\begin{align}
I(t) &=  \int A(t') \, h_\text{RRC}(t - t') \, dt' = \sum_ka_k\ h_{\text{RRC}}(t-k\ T_{\mathrm{sym}}), \\
Q(t) &= \int B(t') \, h_\text{RRC}(t - t') \, dt'= \sum_kb_k\ h_{\text{RRC}}(t-k\ T_{\mathrm{sym}}),
\end{align}
\end{subequations}
and are then modulated onto a carrier frequency $\omega_c$ with a $\pi/2$ phase offset. The resulting transmitted \gls{qam} signal is given by
\begin{equation}
y(t) = \left[ I(t) + i Q(t) \right] e^{i \omega_c t} = \sum_{k} (a_k + i b_k) \, h_{\text{RRC}}(t - k\ T_{\mathrm{sym}}) \, e^{i\omega_c t}.
\end{equation}

For the purposes of analysis and simulation, it is useful to characterize the statistical properties of the QAM signal $s(t)$. Let the transmitted signal be expressed as $s(t)$ and $a_k$, $b_k$ are i.i.d.\ random variables uniformly distributed over $\{0,1,\dots,\sqrt{M}-1\}$. Defining $c_k = a_k + \ii b_k$, the expression simplifies to
\begin{align}
    y(t) = e^{i \omega_c t} \sum_{k} c_k \, h_{\mathrm{RRC}}(t-k\,T_{\mathrm{sym}}).
\end{align}

In the interaction picture, the rapidly oscillating carrier factor
$e^{\ii\omega_c t}$ is removed from the coupled positive-$P$ propagation
equations. Let
\begin{equation}
\widetilde{s}(t)
=\sum_k c_k\, h_{\mathrm{RRC}}(t-k\,T_{\mathrm{sym}})
\end{equation}
denote the resulting unnormalized baseband envelope. In order to scale $\tilde s(t)$, such that the maximum value is approximately 1, we define a scaling factor
\begin{equation}
    C = \frac{1}{\sqrt{2}(\sqrt{M}-1)}, \qquad s(t) = C \tilde s(t).
\end{equation}

The scaled waveform, therefore, satisfies
\begin{equation}
|s(t)|  \lesssim 1. 
\end{equation}

\begin{figure*}[ht]
     \centering
     \includegraphics[width=0.95\linewidth]{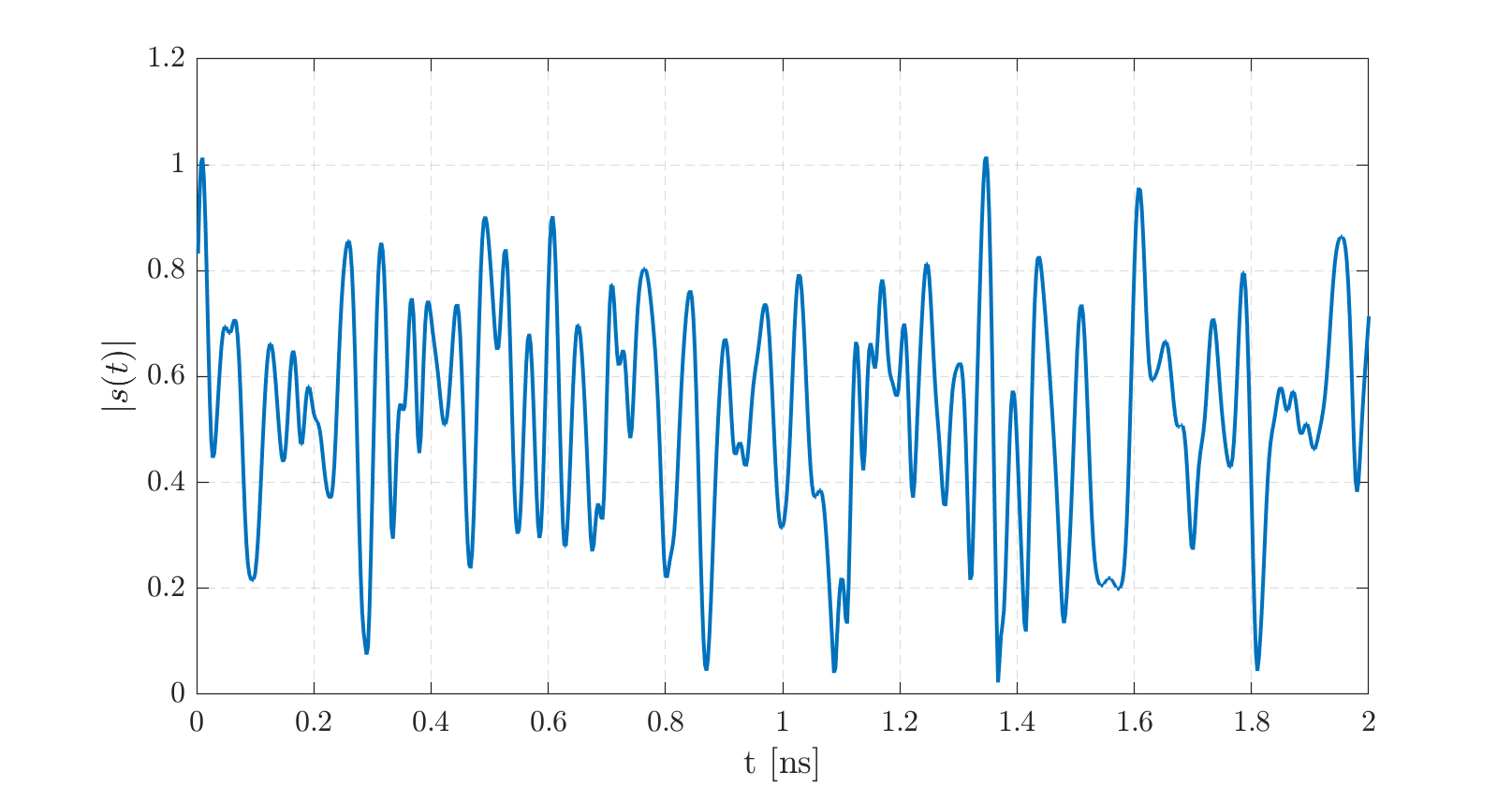}
    \caption{Representative $2\,\mathrm{ns}$ segment of the magnitude of the normalized RRC-shaped $64$-QAM baseband envelope $s(t)$ used for the classical channel. The waveform is generated at the nominal symbol rate $f_{\mathrm{sym}}=50,\mathrm{Gbaud}$ with roll-off factor $\beta=0.2$. Small excursions above unity result from temporal overlap between adjacent RRC-shaped symbols.}
\label{fig:QAM_waveform}
 \end{figure*}

Because $|e^{\ii\omega_c t}|=1$, removing the carrier does not affect this
scaling. Consequently, the power scale applied to $s(t)$ directly
sets the peak classical launch power.

The peak power can be converted into the average power of the \gls{qam} signal by multiplying with $C_{\mathrm{avg}}$. Over the simulated time interval $\mathcal{T}$ of duration $T_{\mathrm{obs}}$,
\begin{equation}
C^{\mathrm{(avg)}}=
\frac{1}{T_{\mathrm{obs}}}
\int_{\mathcal{T}}
|{s}(t)|^2 \,\mathrm{d}t.
\end{equation}

Therefore, if $P_2$ is the peak power of the classical signal, then the average power of the classical signal will be  $P_2^{\mathrm{(avg)}} = C^{\mathrm{(avg)}} P_2$.

In the interaction picture, the rapidly oscillating carrier factor $e^{\ii\omega_c t}$ is removed from the coupled positive-$P$ propagation equations. The resulting unnormalized baseband envelope is
\begin{align}
\widetilde{s}(t)
&=
\sum_k c_k h_{\mathrm{RRC}}(t-kT_{\mathrm{s}}).
\end{align}

For the square $M$-QAM constellation used here, the maximum magnitude of a constellation point is $\sqrt{2}(\sqrt{M}-1)$. We therefore define
\begin{align}
C_{\mathrm{QAM}}
&=
\frac{1}{\sqrt{2}(\sqrt{M}-1)},
\qquad
s(t)=C_{\mathrm{QAM}}\widetilde{s}(t).
\end{align}
This normalization sets the outermost QAM constellation points to unit magnitude. Although overlap between neighboring RRC pulses can produce small overshoots, the scaled waveform remains approximately bounded by
\begin{align}
|s(t)|
&\lesssim 1.
\end{align}
Because $|e^{\ii\omega_c t}|=1$, removing the carrier does not affect this scaling. Consequently, $P_2$ sets the peak classical launch-power scale. A representative segment of the normalized RRC-shaped QAM waveform is shown in Fig.~\ref{fig:QAM_waveform}.

Over the simulated interval $\mathcal{T}$ of duration $T_{\mathrm{obs}}$, the average-to-nominal power ratio is
\begin{align}
C_{\mathrm{avg}}
&=
\frac{1}{T_{\mathrm{obs}}}
\int_{\mathcal{T}}
|s(t)|^2\,\mathrm{d}t.
\label{eq:C_avg}
\end{align}

For the QAM symbol sequence and observation window used in the simulations, numerical evaluation of Eq.~\eqref{eq:C_avg} gives $C_{\mathrm{avg}}=0.347$.

The corresponding average classical launch power is therefore
\begin{align}
P_2^{(\mathrm{avg})}
&=
C_{\mathrm{avg}}P_2.
\end{align}

\IfFileExists{ref.bib}{%
\bibliographystyle{IEEEtran}
\bibliography{ref}
}{}